\documentclass{article}

\pdfoutput=1
\usepackage{scrextend}
\usepackage{graphicx}
\usepackage{amssymb}
\usepackage{amstext}
\usepackage{amsmath}
\usepackage{epstopdf}
\usepackage{booktabs}
\usepackage{verbatim}
\usepackage{gensymb}
\usepackage{geometry}
\usepackage{appendix}
\usepackage{lmodern}
\usepackage{framed}

\usepackage{color} 
\usepackage[T1]{fontenc} 
\usepackage{lmodern} 
\usepackage{helvet} 
\usepackage{xcolor,mdframed,caption} 
\usepackage{fancyhdr}
\usepackage{datetime}
\usepackage{setspace}
\usepackage{sectsty}

\definecolor{ornlGreen}{cmyk}{0.96,0.27,1.00,0.15}
\definecolor{ornlGray}{RGB}{106,100,100}

\usepackage{url}

\newgeometry{margin = 1in}
\renewcommand{\frame}[1]{\begin{framed} {\color{black} #1} \end{framed}}

\usepackage{titlesec} 
\sectionfont{\color{ornlGreen}}

\usepackage{ornl}
\newdateformat{dateMMYYYY}{\monthname[\THEMONTH] \THEYEAR}
\newcommand*\Title{Resilience Design Patterns}
\newcommand*\Subtitle{A Structured Approach to Resilience at Extreme Scale}
\newcommand*\ReportID{ORNL/TM-2016/687}
\newcommand*\Date{October 2016}
\newcommand*\Author{Saurabh Hukerikar \& Christian Engelmann}
\newcommand*\DocumentType{ORNL Technical Report}
\newcommand*\DocumentVersion{1.0}
\author{Saurabh Hukerikar\\
        Christian Engelmann\\
        \{hukerikarsr, engelmannc\}@ornl.gov}

\begin{document}
\begin{titlepage}
\maketitle
\end{titlepage}

\begin{Disclaimer}
\end{Disclaimer}

\begin{DocTitlePage}
\end{DocTitlePage}

\begin{Abstract}
Reliability is a serious concern for future extreme-scale high-performance computing (HPC) systems. Projections based on the current generation of HPC systems and technology roadmaps suggest that very high fault rates in future systems. The errors resulting from these faults will propagate and generate various kinds of failures, which may result in outcomes ranging from result corruptions to catastrophic application crashes. Practical limits on power consumption in HPC systems will require future systems to embrace innovative architectures, increasing the levels of hardware and software complexities. 

The resilience challenge for extreme-scale HPC systems requires management of various hardware and software technologies that are capable of handling a broad set of fault models at accelerated fault rates. These techniques must seek to improve resilience at reasonable overheads to power consumption and performance. While the HPC community has developed various solutions, application-level as well as system-based solutions, the solution space of HPC resilience techniques remains fragmented. There are no formal methods and metrics to investigate and evaluate resilience holistically in HPC systems that consider impact scope, handling coverage, and performance \& power efficiency across the system stack. Additionally, few of the current approaches are portable to newer architectures and software ecosystems, which are expected to be deployed on future systems.

In this document, we develop a structured approach to the management of HPC resilience based on the concept of resilience-based design patterns. A design pattern is a general repeatable solution to a commonly occurring problem. We identify the commonly occurring problems and solutions used to deal with faults, errors and failures in HPC systems. The catalog of resilience design patterns provides designers with reusable design elements. We define a design framework that enhances our understanding of the important constraints and opportunities for solutions deployed at various layers of the system stack. The framework may be used to establish mechanisms and interfaces to coordinate flexible fault management across hardware and software components. The framework also enables optimization of the cost-benefit trade-offs among performance, resilience, and power consumption. The overall goal of this work is to enable a systematic methodology for the design and evaluation of resilience technologies in extreme-scale HPC systems that keep scientific applications running to a correct solution in a timely and cost-efficient manner in spite of frequent faults, errors, and failures of various types.

\end{Abstract}

\tableofcontents
\listoffigures
\listoftables

\ReportContentBegin

\section{Introduction}

High-performance computing systems enable transformative scientific research and discovery in various areas of national importance through computational modeling, simulation, data analysis and prediction. The opportunities to address complex challenges that are important for national security, environmental issues, as well as to drive fundamental scientific research are the key motivators behind the HPC community's drive towards extreme-scale high-performance computing systems. Future extreme-scale systems will enable computing at scales in the hundreds of petaflops, exaflops, and beyond, which will provide the computing capability for rapid design and prototyping and big data analysis in a variety of scientific and engineering disciplines. However, to build and effectively operate extreme-scale HPC systems, there are several key challenges, including the management of power, massive concurrency and resilience to the occurrence of faults or defects in system components \cite{Dongarra:2011:IES}.  

In the pursuit of greater computational capabilities, the architectures of future extreme-scale HPC systems are expected to change radically. Traditional HPC system design methodologies have not had to account for power constraints, or parallelism on the level designers must contemplate for future extreme-scale systems \cite{Shalf:2011}. The evolution in the architectures will require changes to the programming models and the software environment to ensure application scalability. Many of the innovations in the architectures are expected to be driven by the continued scaling of transistors made possible by Moore's Law. However, the reliability of these systems will be threatened by a decrease in individual device reliability due to manufacturing defects prevalent at deeply scaled technology nodes, device aging related effects \cite{Borkar:2005}. Additionally, the chips built using these devices will be increasingly susceptible to errors due to the effects of operational and environmental conditions on the reduced noise margins arising from the near-threshold voltage (NTV) operation \cite{Dreslinski:2010} to meet the limits on power consumption. These effects are expected to increase the rate of transient and hard errors, such that scientific applications running on these future systems will no longer be able to assume correct behavior of the underlying machine.  

Managing the resilience of future extreme-scale systems is a multidimensional challenge. As these systems approach exaflops scale, the frequency of faults and errors in these systems will render many of the existing resilience techniques ineffective. Newer modes of failures due to faults and errors that will only emerge in advanced process technologies and complex system architectures will require novel resilience solutions, as well as existing techniques to adapt their respective implementations. Additionally, HPC resilience methodologies, both hardware and software, must optimize for some combination of performance, power consumption and cost while providing effective protection against faults, errors and failures. The HPC research community has developed a number of hardware and software resilience technologies, but there are no formalized, comprehensive methods to investigate and evaluate the protection coverage and efficiency of resilience solutions. The development of HPC resilience solutions no longer relies on the invention of novel methodologies for dealing with extreme rates and a variety of fault types that may occur; rather, it based on the selection of the most appropriate solutions among the well-understood resilience techniques and adapting them to the design concerns and constraints of future extreme-scale systems. Therefore, the designers of HPC hardware and software components have a compelling need for a systematic methodology for designing resilience solutions for HPC systems and their applications. 

In this work, we develop a structured approach for resilient extreme-scale high-performance computing systems. In general, resilience solutions provide techniques to manage faults and their consequences in a system through the detection of errors in the system, and providing the means that ensure the limiting their propagation, and the recovery of the system from the error/failure, or the masking of error/failure. 
Using the concept of design patterns for resilience, we identify and evaluate repeatedly occurring resilience problems and solutions throughout the hardware/software stack and across various system components. These patterns, which are based on best practices for HPC resilience, are abstracted and codified into a catalog of design patterns. The patterns in this document present solutions for each of the three aspects of resilience, which are the detection, containment, and recovery from an error or a failure. The resilience design patterns describe solutions that are free of implementation details, and these have the potential to shape the design of HPC applications' algorithms, numerical libraries, system software, and hardware architectures, as well as the interfaces between layers of system abstraction. 

In addition to the catalog of resilience design patterns, which provides HPC designers with a collection of reusable design elements, we define a discipline that enables designers to combine an essential set of design patterns into productive, efficient resilience solutions. We define a layered hierarchy of these resilience design patterns, which facilitates the design process of combining the various patterns in the catalog. The classification scheme provides guidelines for designers to distinctly solve problems of detection, containment and mitigation/recovery, to stitch these patterns together, and refine the design of the overall system based on the roles of the individual patterns, and how they interact. We define a framework based on \textit{design spaces} that guides hardware and software designers and architects, as well as application developers, in navigating the complexities of developing effective resilience solutions within the constraints of hardware and software implementation challenges, performance and power considerations. The construction of resilience solutions using the design patterns and using the design spaces framework makes the design choices explicit and the critical issues addressed by a solution clear. The structured, pattern-based approach enables:   

\begin{itemize}
\item Development of resilience solutions with a clear understanding of their protection coverage and efficiency  
\item Evaluation and comparison of alternative resilience solutions through qualitative and quantitative evaluation of the coverage and handling efficiency of each solution
\item Design portable resilience solutions that are effective on different systems, since each HPC system architecture has different resilience features and solutions are not universally applicable 
\end{itemize}

The rest of this document is organized as follows:
\begin{itemize}
\item Section \ref{sec:Background} provides a summary of the terminology used in fault tolerance and the basic concepts of resilience to enable HPC designers, as well as system operators and users to understand the essence of the resilience patterns and use them in their designs, whether hardware or software.   
\item Section \ref{sec:HPC-Resilience-Challenges} describes the challenges in managing the resilience of future extreme-scale HPC systems. 
\item Section \ref{sec:HPC-Resilience-Survey} surveys the various HPC resilience solutions, including those used in production HPC systems, as well as research proposals. The aim of this section is to provide a comprehensive overview of the various HPC resilience techniques. 
\item Section \ref{sec:Patterns-Concept} introduces the design pattern concept and discusses the potential for capturing the HPC 
resilience techniques in the form of patterns.  
\item Section \ref{sec:Resilience-Patterns-Classification} describes a classification scheme to organize the various resilience techniques in a layered hierarchy to enable designers to understand the capabilities of each resilience solution. 
\item Section \ref{sec:Resilience-Pattern-Catalog} presents the catalog of resilience design patterns that capture well-understood HPC resilience techniques for error detection, recovery and masking in a structured format. 
\item Section \ref{sec:Building-Resilience-Solutions} presents a structured methodology to use the resilience design patterns for the construction of effective and efficient resilience solutions. The design framework forms the basis for HPC designers and programmers to use patterns from the catalog to develop complete resilience solutions with specific properties.    
\item Sections \ref{sec:CaseStudy-CR}, \ref{sec:CaseStudy-ProcessMigration}, \ref{sec:CaseStudy_ECC-ABFT} present case studies that demonstrate how these patterns may be used to understand and evaluate existing resilience solutions, as well as develop new solutions using the pattern-based design framework.  
\end{itemize}

\clearpage

\section{Resilience Terminology and Concepts}
\label{sec:Background}

The terminology is largely based on prior work on establishing agreed upon definitions and metrics for HPC Reliability, Availability \& Servicability (RAS)~\cite{Snir:2014, Koren:2007, Vargas:2000, Engelmann:2008:Thesis, Eusgeld:2005, Pham:2007}.

\subsection{Reliability}
Reliability is the property of a system that characterizes its probability to have an error or failure, {\em i.e.}, it provides information about the error- or failure-free time period.

\subsection{Availability}
Availability is the property of a system that defines the proportion of time it provides a correct service, instead of incorrect service.

\subsection{Systems}
\label{section:taxonomy:system}
\begin{itemize}
  \item {\bf System:} An entity that interacts with other entities.
  \item {\bf Component:} A system that is part of a larger system.
  \item {\bf State:} A system's information, computation, communication, interconnection, and physical condition.
  \item {\bf Behavior:} What a system does to implement its function, described by a series of states.
  \item {\bf Service:} A system's externally perceived behavior.
  \item {\bf Functional specification:} The description of system functionality, defining the threshold between:
    \begin{itemize}
      \item {\bf Correct service:} The provided service is acceptable, i.e., within the functional specification.
      \item {\bf Incorrect service:} The provided service is unacceptable, i.e., outside the functional specification.
    \end{itemize}
  \item {\bf Life cycle:} A system has life cycle phases in the following order:
  \begin{enumerate}
    \item {\bf Development:} A system is in development, including designed, constructed, deployed and tested.
    \item {\bf Operational:} A system is in operation, providing correct or incorrect service.
    \item {\bf Retired:} A system is not in operation anymore.
  \end{enumerate}
  \item {\bf Operational status:} A system has the following operational states:
    \begin{enumerate}
      \item {\bf Scheduled service outage:} A system is delivering an incorrect service due to a planned outage.
      \item {\bf Unscheduled service outage:} A system is delivering an incorrect service due to an unplanned outage.
      \item {\bf Service delivery:} A system is delivering a correct service.
    \end{enumerate}
\end{itemize}

The terms \textbf{fault}, \textbf{error} and \textbf{failure} are sometimes used interchangeably. However in fault tolerance literature, these terms are associated with distinct formal concepts which are defined as follows \cite{Avizienis:2004}:

\subsection{Faults}
\textbf{Fault} is an underlying flaw or defect in a system that has potential to cause problems. A fault can be dormant and can have no effect, e.g., incorrect program code that lies outside the execution path. When activated during system operation, a fault leads to an error.  Fault activation may be due to triggers that are internal or external to the system.

\begin{itemize}
  \item {\bf Fault classes:}
    {\em\{benign,dormant,active\}
         \{permanent,transient,intermittent\}
         \{hard,soft\}}
\end{itemize}

These fault classes have the following categories:
\begin{itemize}
  \item {\bf Benign:} An inactive fault that does not become active.
  \item {\bf Dormant:} An inactive fault that does become active at some point in time.
  \item {\bf Active:} A fault that causes an error at the moment of becoming active.
  \item {\bf Permanent:} A fault's presence is continuous in time.
  \item {\bf Transient:} A fault's presence is temporary.
  \item {\bf Intermittent:} A fault's presence is temporary and recurring.
  \item {\bf Hard:} A fault that is systematically reproducible.
  \item {\bf Soft:} A fault that is not systematically reproducible.
\end{itemize}

The following common terms map to these fault classes:
\begin{itemize}
  \item {\bf Latent fault:} Any type of {\em dormant fault}.
  \item {\bf Solid fault:} Any type of {\em hard fault}.
  \item {\bf Elusive fault:} Any type of {\em soft fault}.
\end{itemize}

For example, a radiation-induced bit-flip in memory is a {\em dormant transient soft fault} that becomes an {\em active transient soft fault} when the memory is read. The fault disappears when the memory is written. A radiation-induced bit-flip in memory is a {\em dormant permanent soft fault} if the memory is never written. It becomes an {\em active permanent soft fault} when the memory is read.

\subsection{Errors}
\textbf{Errors} result from the activation of a fault and cause an illegal system state. For e.g., a faulty assignment to a loop counter variable may result in an error characterized by an illegal value for that variable. When such a variable is used for control of a for-loop's execution, it may lead to incorrect program behavior.

The following error classes exist:
\begin{itemize}
  \item {\bf Error classes:}
    {\em\{undetected,detected\}
        \{unmasked,masked\}
        \{hard,soft\}}
\end{itemize}

These error classes have the following categories:
\begin{itemize}
  \item {\bf Undetected:} An error that is not indicated.
  \item {\bf Detected:} An error that is indicated, such as by a message or a signal.
  \item {\bf Unmasked:} An error that is propagating.
  \item {\bf Masked:} An error that is not propagating.
  \item {\bf Hard:} An error caused by a permanent fault.
  \item {\bf Soft:} An error caused by a transient or intermittent fault.
\end{itemize}

The following common terms map to these error classes:
\begin{itemize}
  \item {\bf UE:} Any type of {\em undetected error}.
  \item {\bf Latent error:} Any type of {\em undetected error}.
  \item {\bf Silent error:} Any type of {\em undetected error}.
  \item {\bf SDC:} An {\em undetected unmasked hard} or {\em soft error}.
  \item {\bf DE:} Any type of {\em detected error}.
\end{itemize}

For example, an {\em active transient soft fault}, created by a radiation-induced bit-flip in memory being read, causes an {\em undetected masked soft error}, when the read value is used in a multiplication with another value that happens to be $0$. It causes an {\em undetected unmasked soft error}, or SDC, when the read value is used as an index in a memory address calculation.

A detectable correctable error is often transparently handled by hardware, such as a single bit flip in memory that is protected with single-error correction double-error detection (SECDED) error correcting code (ECC) \cite{Moon:2005}. A detectable uncorrectable error (DUE) typically results in a failure, such as multiple bit flips in the same addressable word that escape SECDED ECC correction, but not detection, and ultimately cause an application abort. An undetectable error may result in silent data corruption (SDC), e.g., an incorrect application output.

\subsection{Failures}
\textbf{Failure} occurs if an error reaches the service interface of a system, resulting in system behavior that is inconsistent with the system's specification. For e.g., a faulty assignment to a pointer variable leads to erroneous accesses to a data structure or buffer overflow, which in turn may cause the program to crash due to an attempt to access an out-of-bound memory location.

The following failure classes exist:
\begin{itemize}
  \item {\bf Failure classes:}
    {\em\{undetected,detected\}
        \{permanent,transient,intermittent\}
        \{complete,partial,Byzantine\}}
\end{itemize}

These failure classes have the following categories:

\begin{itemize}
  \item {\bf Undetected:} A failure that is not indicated.
  \item {\bf Detected:} A failure that is indicated, such as by a message or a signal.
  \item {\bf Permanent:} A failure's presence is continuous in time.
  \item {\bf Transient:} A failure's presence is temporary.
  \item {\bf Intermittent:} A failure's presence is temporary and recurring.
  \item {\bf Complete:} A failure causing a service outage.
  \item {\bf Partial:} A failure causing a degraded service within the functional specification.
  \item {\bf Byzantine:} A failure causing an arbitrary deviation from the functional specification.
\end{itemize}

The following common terms map to these error classes:
\begin{itemize}
  \item {\bf Fail-stop:} An {\em undetected} or {\em detected permanent complete failure}.
\end{itemize}

For example, an {\em active transient soft fault}, created by a radiation-induced bit-flip in memory being read, causes an {\em undetected unmasked soft error}, when the read value is used as an index in a memory address calculation. A memory access violation caused by using a corrupted calculated address results in a {\em detected permanent complete failure}, as the executing process is killed by the operating system (OS), and a message is provided to the user. However, if using the corrupted calculated address results in an incorrect service that is not indicated, such as erroneous output, an {\em undetected intermittent Byzantine failure} occured.

\subsection{The Relationship between Faults, Errors and Failures}
\begin{figure}[tp]
\centering
\includegraphics[width=90mm]{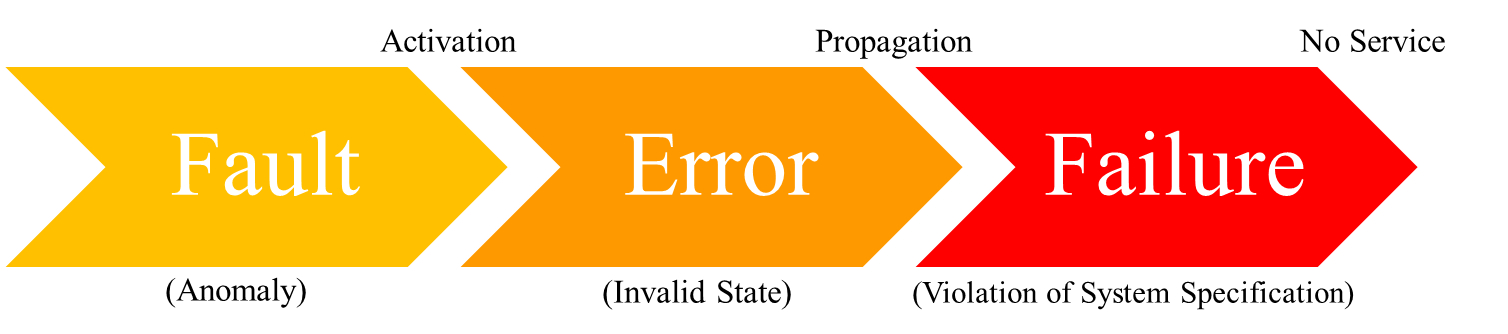}
\caption{Relationship between fault, error and failure}
\label{Fig:Fault-Error-Failure}
\end{figure}

While a fault (e.g., a bug or stuck bit) is the cause of an error, its manifestation as a state change is considered an error (e.g., a bad value or incorrect execution), and the transition to an incorrect service is observed as a failure (e.g., an application abort or system crash) \cite{Snir:2014}. A failure in a HPC system is typically observed through an application abort or a full/partial system outage. There is a causality relationship between fault, error and failure, as shown in Figure \ref{Fig:Fault-Error-Failure}. A \textbf{Fault-error-failure chain} is a DAG representation with faults, errors and failures as vertices. When the system is composed of multiple components, the failure of a single component causes a permanent or transient external fault for the other components that receive service from the failed component. Therefore, errors may be transformed into other errors and propagate through the system, generating further errors. A \textbf{failure cascade} is a failure of component $B$ that cascades to component $C$ if it causes a failure of $C$. For example, a faulty procedure argument leads to erroneous computation and may manifest as a failure in the form of an illegal procedure return value. To the caller of the function, this may activate a chain of errors that propagate until service failure occurs, i.e., a program crash.

\subsection{Resilience Capabilities}
There are three key components to designing a resilience strategy:

\subsubsection{Detection}
Detection entails the discovery of an error in the state of the system, either in the data, or in the instructions. It is typically accomplished with the help of redundancy; the extra information enables the verification of correct values.

Errors are detected by identifying the corresponding state change. Failures are detected by identifying the corresponding transition to an incorrect service. An error or a failure is indicated by a detector. This detector can fail as well.

The following detection classes exist:
\begin{itemize}
  \item {\bf Detection classes:}
    {\em\{true,false\}
         \{positive,negative\}}
\end{itemize}

These detection classes have the following categories:
\begin{itemize}
  \item {\bf True:} A correct detection.
  \item {\bf False:} An incorrect detection.
  \item {\bf Positive:} An indication, such as a message or a signal.
  \item {\bf Negative:} No indication.
\end{itemize}

\subsubsection{Containment}
A containment capability enables limiting the effects of an error from propagating. Containment is achieved by reasoning about the modularity of components or sub-systems that make up the system. In terms of resilience of the system, a containment module is an unit that fails independently of other units and it is also the unit of repair or replacement.

\subsubsection{Masking}
Masking enatils recovery or mitigation, which ensures correct operation despite the occurrence of an error. Masking is usually accomplished by providing additional redundant in order to discover correct, or at least acceptably close, values of the erroneous state. When the masking involves the change of incorrect state into correct state, it is called error correction.

A detected error may be masked by an error correction method, such as using ECC. The following error classes exist that are equivalent to already defined error classes:
\begin{itemize}
  \item {\bf Error classes:}
    {\em\{uncorrected,corrected\}}
\end{itemize}

These error classes have the following categories:
\begin{itemize}
  \item {\bf Uncorrected:} An error that is not corrected, i.e., an
        {\em undetected} or {\em detected unmasked error}.
  \item {\bf Corrected:} An error that is detected and corrected, i.e., a
        {\em detected masked error}.
\end{itemize}

The following common terms map to these error classes:
\begin{itemize}
  \item {\bf DUE:} An {\em uncorrected error}, i.e., a {\em detected unmasked error}.
  \item {\bf DCE:} A {\em corrected error}, i.e., a {\em detected masked error}.
\end{itemize}

In practice, a resilience mechanism may merge the implementation of two or even all three of the capabilities to provide a complete solution.

\subsection{Resilience Metrics}

\subsubsection{Reliability Metrics}
The following reliability metrics exist:
\begin{itemize}
  \item {\bf Error} or {\bf failure reliability:} A system's probability not to have an error or failure during $0 \leq t$, $R(t)$.
  \item {\bf Error} or {\bf failure distribution:} A system's probability to have an error or failure during $0 \leq t$, $F(t)$.
  \item {\bf PDF:} The relative likelihood of an error or failure, $f(t)$.
  \item {\bf Error} or {\bf failure rate:} A system's error or failure frequency, $\lambda(t)$.
  \item {\bf MTTE:} A system's expected time to error, $MTTE$.
  \item {\bf MTTF:} A system's expected time to failure, $MTTF$.
  \item {\bf FIT rate:} A system's number of expected failures in $10^{9}$ hours of operation, $FIT$.
  \item {\bf Serial reliability:} The reliability of a system with $n$ dependent components, $R(n,t)_{s}$.
  \item {\bf Parallel reliability:} The reliability of a system with $n$ redundant components, $R(n,t)_{p}$.
  \item {\bf Identical serial reliability:} The serial reliability with $n$ identical components, $R(n,t)_{is}$.
  \item {\bf Identical parallel reliability:} The parallel reliability with $n$ identical components, $R(n,t)_{ip}$.
\end{itemize}
\vspace{-0.5\baselineskip}
\begin{align}
  \label{equation:taxonomy:reliability:reliability}
  R(t)            &= 1 - F(t)
                   = \int^{\infty}_{t} f(t) dt\\[0.5\baselineskip]
  \label{equation:taxonomy:reliability:failure}
  F(t)            &= 1 - R(t)
                   = \int^{t}_{0} f(t) dt\\[0.5\baselineskip]
  \label{equation:taxonomy:reliability:rate}
  \lambda(t)      &= \frac{f(t)}{R(t)}\\[0.5\baselineskip]
  \label{equation:taxonomy:reliability:mttf}
  MTTE\ or\ MTTF  &= \int^{\infty}_{0} R(t) dt
\end{align}
\begin{align}
  \label{equation:taxonomy:reliability:fit}
  FIT             &= \frac{10^{9}}{MTTF}\\[0.5\baselineskip]
  \label{equation:taxonomy:reliability:reliability:series}
  R(n,t)_{s}      &= \prod_{i=1}^{n}R_{i}(t)\\[0.5\baselineskip]
  \label{equation:taxonomy:reliability:reliability:parallel}
  R(n,t)_{p}      &= 1 - \prod_{i=1}^{n}(1-R_{i}(t))\\[0.5\baselineskip]
  \label{equation:taxonomy:reliability:reliability:series:identical}
  R(n,t)_{is}     &= R(t)^{n}\\[0.5\baselineskip]
  \label{equation:taxonomy:reliability:reliability:parallel:identical}
  R(n,t)_{ip}     &= 1 - (1-R(t))^{n}
\end{align}

\subsubsection{Availability Metrics}
Availability is the property of a system that defines the proportion of time it provides a correct service, instead of incorrect service. The following availability metrics exist:
\begin{itemize}
  \item {\bf Availability:} A system's proportion of time it provides a correct service, instead of incorrect service, $A$.
  \item {\bf PU:} A system's {\em service delivery} time, $t_{pu}$.
  \item {\bf UD:} A system's {\em unscheduled service outage} time, $t_{ud}$.
  \item {\bf SD:} A system's {\em scheduled service outage} time, $t_{sd}$.
  \item {\bf MTTR:} A system's expected time to repair/replace, $MTTR$.
  \item {\bf MTBF:} A system's expected time between failures, $MTBF$.
  \item {\bf Serial availability:} The availability of a system with $n$ dependent components, $A_{s}$.
  \item {\bf Parallel availability:} The availability of a system with $n$ redundant components, $A_{p}$.
  \item {\bf Identical serial availability:} The serial availability with $n$ identical components, $A_{is}$.
  \item {\bf Identical parallel availability:} The parallel availability with $n$ identical components, $A_{ip}$.
\end{itemize}
\vspace{-0.5\baselineskip}
\begin{align}
  \label{equation:taxonomy:availability:availability:1}
  A            &= \frac{t_{pu}}{t_{pu}+t_{ud}+t_{sd}}\\[0.5\baselineskip]
  \label{equation:taxonomy:availability:availability:2}
               &= \frac{MTTF}{MTTF+MTTR}\\[0.5\baselineskip]
  \label{equation:taxonomy:availability:availability:3}
               &= \frac{MTTF}{MTBF}\\[0.5\baselineskip]
  \label{equation:taxonomy:availability:mtbf}
  MTBF         &= MTTF+MTTR\\[0.5\baselineskip]
  \label{equation:taxonomy:availability:availability:series}
  A_{s}        &= \prod_{i=1}^{n}A_{i}\\[0.5\baselineskip]
  \label{equation:taxonomy:availability:availability:parallel}
  A_{p}        &= 1-\prod_{i=1}^{n}(1-A_{i})\\[0.5\baselineskip]
  \label{equation:taxonomy:availability:availability:series:identical}
  A_{is}       &= A^{n}\\[0.5\baselineskip]
  \label{equation:taxonomy:availability:availability:parallel:identical}
  A_{ip}       &= 1-(1-A)^{n}
\end{align}

A system can also be rated by the number of 9s in its availability figure (Table~\ref{table:taxonomy:availability:nines}). For example, a system with a five-nines availability rating has 99.999\% availability and an annual UD of 5 minutes and 15.4 seconds.
\begin{table}[ht]
  \begin{center}
    \caption{Availability measured by the ``nines''}
    \label{table:taxonomy:availability:nines}
    \begin{tabular}{@{}lll@{}}
      \hline
      9s & Availability & Annual Downtime\\
      \hline
      1  & 90\%       & 36 days,    12   hours   \\
      2  & 99\%       & 87 hours,   36   minutes \\
      3  & 99.9\%     &  8 hours,   45.6 minutes \\
      4  & 99.99\%    & 52 minutes, 33.6 seconds \\
      5  & 99.999\%   &  5 minutes, 15.4 seconds \\
      6  & 99.9999\%  & 31.5 seconds             \\
      \hline
    \end{tabular}
  \end{center}
\end{table}

\subsubsection{Error and Failure Detection Metrics}

\begin{itemize}
  \item {\bf Precision:} The fraction of indicated errors or failures that were actual errors or failures.
  \item {\bf Recall:} The fraction of errors or failures that were detected and indicated.
\end{itemize}
\begin{align}
  \label{equation:taxonomy:detection:precision:1}
  Precision &= \frac{True\ Positives}
                    {True\ Positives + False\ Positives}
             = \frac{True\ Positives}
                    {Indicated\ Errors\ or\ Failures}\\[0.5\baselineskip]
  \label{equation:taxonomy:detection:precision:2}
            &= 1 - \frac{False\ Positives}
                        {True\ Positives + False\ Positives}
             = 1 - \frac{False\ Positives}
                        {Indicated\ Errors\ or\ Failures}\\[0.5\baselineskip]
  \label{equation:taxonomy:detection:recall:1}
  Recall    &= \frac{True\ Positives}
                    {True\ Positives + False\ Negatives}
             = \frac{True\ Positives}
                    {Errors\ or\ Failures}\\[0.5\baselineskip]
  \label{equation:taxonomy:detection:recall:2}
            &= 1 - \frac{False\ Negatives}
                        {True\ Positives + False\ Negatives}
             = 1 - \frac{False\ Negatives}{Errors\ or\ Failures}
\end{align}

For example, a {\em true positive} detection corresponds to an existing error or failure being indicated, while a {\em false positive detection} corresponds a non-existing error or failure being indicated. A {\em true negative} detection corresponds to a non-existing error or failure not being indicated, while a {\em false negative} detection corresponds to an existing error or failure not being indicated.

\subsubsection{Mean Time to Failure}
Resilience is measured by vendors and operators from the system perspective, e.g., by system mean-time to failure (SMTTF) and system mean-time to repair (SMTTR). Users measure resilience from the application perspective, e.g., by application mean-time to failure (AMTTF) and application mean-time to repair (AMTTR) \cite{Snir:2014}. Both perspectives are quite different \cite{Daly:2008}. For example, an application abort caused by a main memory DUE does not require the system to recover, i.e., the SMTTR is 0. However, the aborted application needs to recover its lost state after it has been restarted, i.e., the AMTTR may be hours. Conversely, a failure of a parallel file system server may only impact a subset of the running applications, as the other ones access a different server. In this case, the server failure is counted toward the SMTTF, while the AMTTF differs by application.

\clearpage

\section{The Resilience Challenge for Extreme-Scale HPC Systems}
\label{sec:HPC-Resilience-Challenges}

Various studies that analyze faults, errors and failures in HPC systems indicate that faults are not rare events in large-scale systems and that the distribution of failure root cause is dominated by faults that originate in hardware. These may include faults due to radiation-induced effects such as particle strikes from cosmic radiation, circuit aging related effects, and faults due to chip manufacturing defects and design bugs that remain undetected during post-silicon validation and manifest themselves during system operation. With aggressive scaling of CMOS devices, the amount of charge required to upset a gate or memory cell is decreasing with every process shrink. For very fine transistor feature sizes, the lithography used in patterning transistors causes variations in transistor geometries such as line-edge roughness, body thickness variations and random dopant fluctuations. These lead to variations in the electrical behavior of individual transistor devices, and this manifests itself at the circuit-level in the form of variations in circuit delay, power, and robustness \cite{Austin:2008:IEEE}. The challenge of maintaining resilience continues to evolve as process technology continues to shrink and system designers will use components that operate at lower threshold voltages. The shrinking noise margins makes the components inherently less reliable and leads to a greater number of manufacturing defects, as well as device aging-related effects. The use of system-level performance and power modulation techniques, such as dynamic voltage/frequency scaling, also tend to induce higher fault rates. It is expected that future exascale-capability systems will use components that have transistor feature sizes between 5 nm and 7 nm, and that these effects will become more prevalent, thereby causing the system components to be increasingly unreliable \cite{Daly:2012}. The modeling and mitigation of these effects through improved manufacturing processes and circuit-level techniques might prove too difficult or too expensive.

Today's petascale-class HPC systems already employ millions of processor cores and memory chips to drive HPC application performance. The recent trends in system architectures suggest that future exascale-class HPC systems will be built from hundreds of millions of components organized in complex hierarchies. However, with the growing number of components, the overall reliability of the system decreases proportionally. If \textit{p} is the probability of failure of an individual component and the system consists of \textit{N} components, the probability that the complete system works is \textit{(1 - p)$^N$} when the component failures are independent. It may therefore be expected that some part of an exascale class supercomputing system will always be experiencing failures or operating in a degraded state. The drop in MTTF of the system is expected to be dramatic based on the projected system features \cite{ExascaleTechStudyReport}. In future exascale-class systems, the unreliability of chips due to transistor scaling issues will be amplified by the large number of components. For long running scientific simulations and analysis applications that will run on these systems, the accelerated rates of system failures will mean that their executions will often terminate abnormally, or in many cases, complete with incorrect results. Finding solutions to these challenges will therefore require a concerted and collaborative effort on the part of all the layers of the system stack.

Resilience is an approach to fault tolerance for high-end computing (HEC) systems that seeks to keep the application workloads running to correct solutions in a timely and efficient manner in spite of frequent errors \cite{Debardeleben:2009}. The emphasis is on the application's outcome and the reliability of application level information in place of or even at the expense of reliability of the system. Resilience technologies in HPC embrace the fact that the underlying fabric of hardware and system software will be unreliable and seek to enable effective and resource efficient use of the platform in the presence of system degradations and failures \cite{Daly:2012}. A complete resilience solution consists of detection, containment and mitigation strategies. 

Performance, resilience, and power consumption are interdependent key system design factors. An increase in resilience (e.g., though redundancy) can result in higher performance (as less work is wasted) and in higher power consumption (as more hardware is being used). Similarly, a decrease in power consumption (e.g., through NTV operation) can result in lower resilience (due to higher soft error vulnerability) and lower performance (due to lower clock frequencies and more wasted work). The performance, resilience, and power consumption cost/benefit trade-off between different resilience solutions depends on individual system and application properties. Understanding this trade-off at system design time is a complex problem due to uncertainties in future system hardware and software reliability. It is also difficult due to a needed comprehensive methodology for design space exploration that accounts for performance, resilience, and power consumption aspects across the stack and the system. Similarly, runtime adaptation to changing resilience demands, while staying within a fixed power budget and achieving maximum performance, is currently limited to checkpoint placement strategies. While resilience technologies seek to provide efficient and correct operation despite the frequent occurrence of faults and defects in components that lead to errors and failures in HPC systems, there is no methodology for optimizing the trade-off, at design time or runtime, between the key system design factors: performance, resilience, and power consumption.

\clearpage

\section{Survey of HPC Resilience Techniques}
\label{sec:HPC-Resilience-Survey}

This section surveys various fault-tolerance techniques used in practical computing systems, as well as research proposals.  

\subsection{Hardware-based Techniques}
Dual-modular redundancy (DMR) and triple-modular redundancy (TMR) approaches have been successfully used in mission-critical systems using hardware-based replication. Examples of fault-tolerant servers include the Tandem Non-Stop \cite{McEvoy:1981:ACM} and the HP NonStop \cite{Bernick:2005:DSN} that use two redundant processors running in locked step. The IBM G5 \cite{Slegel:1999:Micro} employs two fully duplicated lock-step pipelines to enable low-latency detection and rapid recovery. While these solutions are transparent to the supervisor software and application programmer, they require specialized hardware.

Error correction codes (ECC) use a flavor of redundancy in memory structures that typically add additional bits to enable detection and correction of errors. Single bit-error correction and double bit-error detection (SECDED) is the most widely used variant of ECC, while researchers have also explored Bose-Chaudhuri-Hocquenghem (BCH) and double-bit error correction and triple-bit error detection (DECTED) \cite{Naseer:2008:ESSCIRC} for multi-bit detection and correction. Chipkill \cite{IBMChipKill:1997:Whitepaper} is a stronger memory protection scheme that is widely used in production HPC systems. It accommodates single DRAM memory chip failure as well as multi-bit errors from any portion of a single memory chip by interleaving bit error-correcting codes across multiple memory chips.

HPC vendors have also developed a number of hardware resilience technologies, including: SECDED ECC for main memory, caches, registers and architectural state, as well as, Chipkill \cite{Chipkill:Whitepaper:1997} for main memory, redundant power supplies and voltage regulators \cite{Cray:XC40Spec:2014}, and reliability, availability and serviceability management systems for monitoring and control.

\subsection{Software-implemented Techniques}
Software-based redundancy promises to offer more flexibility and tends to be less expensive in terms of silicon area as well as chip development and verification costs; it also eliminates the need for modifications of architectural specifications. 

\subsubsection{Operating System \& Runtime-based Solutions}
The most widely used strategies in production HPC systems are predominantly based on checkpoint and restart (C/R). In general, C/R approaches are based on the concept of capturing the state of the application at key points of the execution, which is then saved to persistent storage. Upon detection of a failure, the application state is restored from the latest disk committed checkpoint, and execution resumes from that point.  The Condor standalone checkpoint library \cite{Litzkow:1992:Condor} was developed to provide checkpointing for UNIX processes, while the Berkeley Labs C/R library \cite{BLCR:2002:LBNL} was developed as an extension to the Linux OS. The libckpt \cite{Plank:1995:Libckpt} provided similar OS-level process checkpointing, albeit based on programmer annotations.

In the context of parallel distributed computing systems, checkpointing requires global coordination, i.e., all processes on all nodes are paused until all messages in-flight and those in-queue are delivered, at which point all the processes' address spaces, register states, etc., are written to stable storage, generally a parallel file system, through dedicated I/O nodes.  The significant challenge in these efforts is the coordination among processes so that later recovery restores the system to a consistent state. These approaches typically launch daemons on every node that form and maintain communication groups that allow tracking and managing recovery by maintaining the configuration of the communication system. The failure of any given node in the group is handled by restarting the failed process on a different node, by restructuring the computation, or through transparent migration to another node \cite{Agbaria:1999:HPDC} \cite{Casas:1995:PVM} \cite{Leon:1993:FailSafePVM}.

Much work has also been done to optimize the process of C/R. A two-level recovery scheme proposed optimization of the recovery process for more probable failures, so that these incur a lower performance overhead while the less probable failures incur a higher overhead \cite{Vaidya:1995:CTD}. The scalable checkpoint/restart (SCR) library \cite{Mohror:2013:SCR} proposes multilevel checkpointing where checkpoints are written to storage that use RAM, flash, or local disk drive, in addition to the parallel file system, to achieve much higher I/O bandwidth. Oliner {\em et al.} propose an opportunistic checkpointing scheme that writes checkpoints that are predicted to be useful - for example, when a failure in the near future is likely \cite{Oliner:2006:ICS}. Incremental checkpointing dynamically identifies the changed blocks of memory since the last checkpoint through a hash function \cite{Agarwal:2004:ICS} in order to limit the amount of state required to be captured per checkpoint. Data aggregation and compression also help reduce the bandwidth requirements when committing the checkpoint to disk \cite{Islam:2012:SCR}. Plank {\em et al.} eliminate the overhead of writing checkpoints to disk altogether with a diskless in-memory checkpointing approach \cite{Plank:1998:IPDPS}.

Process-level redundancy (PLR) \cite{Shye:2009:IEEE} creates a set of redundant application processes whose output values are compared. The scheduling of the redundant processes is left to the operating system (OS). The RedThreads API \cite{Hukerikar:2016:IJPP} provides directives that support error detection and correction semantics through the adaptive use of redundant multithreading.  

\subsubsection{Message Passing Library-based Solutions}
In general, automatic application-oblivious checkpointing approaches suffer from scaling issues due to the considerable I/O bandwidth for writing to persistent storage. Also, practical implementations tend to be fragile \cite{ExascaleResilienceStudy}. Therefore, several MPI libraries have been enabled with capabilities for C/R \cite{Li:1997:PRDC}. The CoCheck MPI \cite{Stellner:1996:PP}, based on the Condor library, uses synchronous checkpointing in which all MPI processes commit their message queues to disk to prevent messages in flight from getting lost. The FT-MPI \cite{Fagg:2000:FTMPI}, Open MPI \cite{Hursey:2007:OpenMPI}, MPICH-V \cite{Bosilca:2002:MPICHV} and LAM/MPI \cite{Sankaran:2003:LAMMPI} implementations followed suit by incorporating similar capabilities for C/R. In these implementations, the application developers do not need to concern themselves with failure handling; the failure detection and application recovery are handled transparently by the MPI library, in collaboration with the OS.

The process-level redundancy approach has also been evaluated in the context of a MPI library implementation \cite{Ferreira:2011:SC}, where each MPI rank in the application is replicated and the replica takes the place of a failed rank, allowing the application to continue. The RedMPI library \cite{Engelmann:2011} \cite{Fiala:2012} replicates MPI tasks and compares the received  messages between the replicas in order to detect corruptions in the communication data. Studies have also proposed the use of proactive fault tolerance in MPI \cite{Nagarajan:2007} \cite{Wang:2008}. However, with the growing complexity of long running scientific applications, complete multi-modular redundancy, whether through hardware or software-based approaches, will incur exorbitant overhead to costs, performance and energy, and is not a scalable solution to be widely used in future exascale-class HPC systems.


\subsubsection{Compiler-based Solutions}
SWIFT \cite{Reis:2005:CGO} is a compiler-based transformation which duplicates all program instructions and inserts comparison instructions during code generation so that the duplicated instructions fill the scheduling slack. The DAFT \cite{Zhang:2010:PACT} approach uses a compiler transformation that duplicates the entire program in a redundant thread that trails the main thread and inserts instructions for error checking. The SRMT \cite{Wang:2007:CGO} uses compiler analysis to generate redundant threads mapped to different cores in a chip multi-processor and optimizes performance by minimizing data communication between the main thread and trailing redundant thread. Similarly, EDDI \cite{Oh:2002} duplicates all instructions and inserts ``compare" instructions to validate the program correctness at appropriate locations in the program code. The ROSE::FTTransform \cite{Lidman:2012} applies source-to-source translation to duplicate individual source-level statements to detect transient processor faults.

\subsubsection{Programming Model Techniques}
Most programming model approaches advocate a collaborative management of the reliability requirements of applications through a programmer interface in conjunction with compiler transformations, a runtime framework and/or library support. Each approach requires different levels of programmer involvement, which has an impact on amount of effort to re-factor the application code, as well as on the portability of the application code to different platforms.

HPC programs usually deploy a large number of nodes to implement a single computation and use MPI with a flat model of message exchange in which any node can communicate with another. Every node that participates in a computation acquires dependencies on the states of the other nodes. Therefore, the failure of a single node results in the failure of the entire computation since the message passing model lacks well-defined failure containment capabilities \cite{ExascaleResilienceStudy}. User-level failure mitigation (ULFM) \cite{Bland:2013:IJHPCA} extends MPI by encouraging programmer involvement in the failure detection and recovery by providing a fault-tolerance API for MPI programs. The error handling of the communicator has changed from MPI\allowbreak\_ERRORS\allowbreak\_ARE\allowbreak\_FATAL to MPI\allowbreak\_ERRORS\allowbreak\_RETURN so that error recovery may be handled by the user. The proposed API includes MPI\allowbreak\_COMM\allowbreak\_REVOKE, MPI\allowbreak\_COMM\allowbreak\_SHRINK to enable reconstruction of the MPI communicator after process failure and the MPI\allowbreak\_COMM\allowbreak\_AGREE as a consistency check to detect failures when the programmer deems such a sanity check necessary in the application code.

The abstraction of the \emph{transaction} has also been proposed to capture a programmer's fault-tolerance knowledge. This entails division of the application code into blocks of code whose results are checked for correctness before proceeding.
If the code block execution's correctness criteria are not met, the results are discarded and the block can be re-executed. Such an approach was explored for HPC applications through a programming construct called \textit{Containment Domains} by Sullivan {\em et al.} \cite{Chung:2011:SC} which is based on weak transactional semantics. It enforces the check for correctness of the data value generated within the containment domain before it is communicated to other domains. These containment domains can be hierarchical and provide the means to locally recover from an error within that domain. A compiler technique that, through static analysis, discovers regions that can be freely re-executed without checkpointed state or side-effects, called idempotent regions, was proposed by de Kruijf {\em et al.} \cite{deKruijf:2012:PLDI}. Their original proposal \cite{deKruijf:2010:ISCA}, however, was based on language-level support for C/C++ that allowed the application developer to define idempotent regions through specification of \emph{relax} blocks and \emph{recover} blocks that perform recovery when a fault occurs.
The FaultTM scheme adapts the concept of hardware-based transactional memory where atomicity of computation is guaranteed. The approach requires an application programmer to define \emph{vulnerable} sections of code. For such sections, a backup thread is created. The original and the backup thread are executed as an atomic transaction, and their respective committed result values are compared \cite{Yalcin:inria:2010:PESPMA}.

Complementary to approaches that focus on resiliency of computational blocks, the Global View Resilience (GVR) project \cite{Fujita:2013:ASPLOS} concentrates on application data and guarantees resilience through multiple snapshot versions of the data whose creation is controlled by the programmer through application annotations. Bridges {\em et al.} \cite{Bridges:2011:Resilience} proposed a {\tt malloc\_failable} that uses a callback mechanism to handle memory failures on dynamically allocated memory, so that the application programmer can specify recovery actions. The Global Arrays implementation of the Partitioned Global Address Space (PGAS) model presents a global view of multidimensional arrays that are physically distributed among the memories of processes. Through a set of library API for checkpoint and restart with bindings for C/C++/FORTRAN, the application programmer can create checkpoints of array structures. The library guarantees that updates to the global shared data are fully completed and any partial updates are prevented or undone \cite{Dinan:2010:CCGrid}. Rolex \cite{Hukerikar:2016} provides various resilience semantics for error tolerance and amelioration through language-based extensions that enable these capabilities to be embedded within standard C/C++ programs. 

\subsubsection{Algorithm-Based Fault Tolerance}
Algorithm-based fault tolerance (ABFT) schemes encode the application data to detect and correct errors, e.g., the use of checksums on dense matrix structures. The algorithms are modified to operate on the encoded data structures. ABFT was shown to be an effective method for application-layer detection and correction by Huang and Abraham \cite{Huang:1984:IEEE} for a range of basic matrix operations including addition, multiplication, scalar product, transposition. Such techniques were also proven effective for LU factorization \cite{Davies:2011:ICS}, Cholesky factorization \cite{Hakkarinen:2010:IPDPS} and QR factorization \cite{Jou:1984:SPIE}. Several papers propose improvements for better scalability in the context of parallel systems, that provide better error detection and correction coverage with lower application overheads \cite{Rexford:1992:ISCAS} \cite{Plank:1995:FTCS} \cite{Roy-Chowdhury:1996:IEEE}. \nocite{Du2012:PPoPP}
The checksum-based detection and correction methods tend to incur very high overheads to performance in sparse matrix-based applications. Sloan {\em et al.}  \cite{Sloan:2012:DSN} have proposed techniques for fault detection that employ approximate random checking and approximate clustered checking by leveraging the diagonal, banded diagonal, and block diagonal structures of sparse problems. Algorithm-based recovery for sparse matrix problems has been demonstrated through error localization and re-computation \cite{Sloan:2013:DSN} \cite{Chen:2011:HPDC}.

Various studies have evaluated the fault resilience of solvers of linear algebra problems \cite{Bronevetsky:2008}. Iterative methods including Jacobi, Gauss-Seidel and its variants, the conjugate gradient, the preconditioned conjugate gradient, and the multi-grid begin with an initial guess of the solution and iteratively approach a solution by reducing the error in the current guess of the answer until a convergence criterion is satisfied. Such algorithms have proved to be tolerant to errors, on a limited basis, since the calculations typically require a larger number of iterations to converge, based on magnitude of the perturbation, but eventual convergence to a correct solution is possible. Algorithm-based error detection in the multigrid method shown by Mishra {\em et al.} \cite{Mishra:2003:IEEE}, uses invariants that enable checking for errors in the relaxation, restriction and the interpolation operators.

For fast Fourier transform (FFT) algorithms, an error-detection technique called the sum-of-squares (SOS) was presented by Reddy {\em et al.} \cite{Reddy:1990:IEEE}. This method is effective for a broader class of problems called orthogonal transforms and therefore applicable to QR factorization, singular-value decomposition, and least-squares minimization. Error detection in the result of the FFT is possible using weighted checksums on the input and output \cite{Wang:1992:ISCAS}. \nocite{Jou:1988:IEEE}

While the previously discussed methods are primarily for numerical algorithms, fault tolerance for other scientific application areas has also been explored. In molecular dynamics (MD) simulations, the property that pairwise interactions are anti-symmetric (F$_{ij}$ = - F$_{ji}$) may be leveraged to detect errors in the force calculations \cite{Yajnik:1994:ISCAS}. The resilience of the Hartree-Fock algorithm, which is widely used in computational chemistry, can be significantly enhanced through checksum-based detection and correction for the geometry and basis set objects. For the two-electron integrals and Fock matrix elements, knowing their respective value bounds allows for identifying outliers and correcting them with reasonable values from a range of known correct values. The iterative nature of the Hartree-Fock algorithm helps to eliminate the errors introduced by the interpolated values \cite{vanDamm:2013:JCompChem}. The fault-tolerant version of the 3D- protein reconstruction algorithm (FT-COMAR) proposed by Vassura {\em et al.} \cite{Vassura:2008:Bioinformatics} is able to recover from errors in as many as 75\% of the entries of the contact map.

\subsection{Cooperative Hardware/Software Approaches} 
Cross-layer resilience techniques \cite{Mitra:2010:DATE} employ multiple error resilience techniques from different layers of the system stack to collaboratively achieve error resilience. These frameworks combine selective circuit-level hardening and logic-level parity checking with algorithm-based fault tolerance methods to provide resilient operation.

\clearpage

\section{Design Patterns for Resilience}
\label{sec:Patterns-Concept}

\subsection{Introduction to Design Patterns}
A \textit{design pattern} describes a generalizable solution to a recurring problem that occurs within a well-defined context. Patterns are often derived from best practices used by designers and they contain essential elements of the problems and their solutions. They provide designers with a template for how to solve a problem that can be used in many different situations. The patterns may also be used to describe design alternatives to a specific problem.

The original concept of design patterns was developed in the context of civil architecture and engineering problems \cite{Alexander:1977}. The patterns were defined with the goal of identification and cataloging solutions to recurrent problems codify and solutions in the building and planning of neighborhoods, towns and cities, as well as in the construction of individual rooms and buildings. In the domain of software architecture, the intent of design patterns isn't to provide a finished design that may be transformed directly into code. Rather, design patterns are used to enhance the software development process by providing proven development paradigms. With the use of design patterns, there is sufficient flexibility for software developers to adapt their implementation to accommodate any constraints, or issues that may be unique to specific programming paradigms, or the target platform for the software.

In the context of object-oriented programming design patterns provide a catalog of methods for defining class interfaces and inheritance hierarchies, and establish key relationships among them \cite{Gamma:1995}. In many object-oriented systems, reusable patterns of class relationships and communicating objects are used to create flexible, elegant, and ultimately reusable software design. In the pursuit of quality and scalable parallel software, patterns for parallel programming were developed \cite{Mattson:2004} and a pattern language, called Our Pattern Language (OPL) \cite{Mattson:OPL:2009}, was used as the means to systematically describe these parallel computation patterns and use them to architect parallel software. In each of these domains of design and engineering, design patterns capture the essence of effective solutions in a succinct form that may be easily applied in similar form to other contexts and problems.

\subsection{Design Patterns for HPC Resilience Solutions}

Since the early days of computing, designers of computer systems repeatedly used well-known techniques to increase reliability: redundant structures to mask failed components, error-control codes and duplication or triplication with voting to detect or correct information errors, diagnostic techniques to locate failed components, and automatic switchovers to replace failed subsystems \cite{Avizienis:1997}. In general, every resilience solution consists of the following capabilities:
\begin{itemize}
\item \textbf{Detection:}
Detecting the presence of errors or failures in the data or control value is an important aspect of any resilience management strategy. The presence of errors in the system are typically detected using redundant information. 

\item \textbf{Containment:}
When an error is detected in a system, its propagation must be limited. The containment of an error requires specification of well-defined modular structures and interfaces. Containment strategies assist in limiting the impact of errors on other modules of the system. 

\item \textbf{Recovery:}
The recovery aspect of any resilience strategy is needed to ensure that the application outcome is correct in spite of the presence of the error. Recovery may entail preventing faults from resurfacing or eliminating the error completely. Rollback and roll-forward are used to position the system state to a previous or forward known correct state. Alternatively, the error may be compensated through redundancy. Recovery may also include system reconfiguration or reinitialization where the system is reset to a set of known parameters which guarantees correct state.
\end{itemize}

Every resilience strategy must contain these core capabilities. However, many design decisions in HPC resilience are unique and the approach to designing a solution may vary considerably based on the layer of system abstraction and the optimization constraints. Each resilience technique provides different guarantees regarding the properties associated to the system qualities such as the time or the space overhead introduced to the normal execution of the system, the efficiency of the reaction to a failure, the design complexity added to the system. 

\subsection{Anatomy of a Resilience Design Pattern}
\begin{figure} [tp]
\centering
\includegraphics[width=0.5\linewidth]{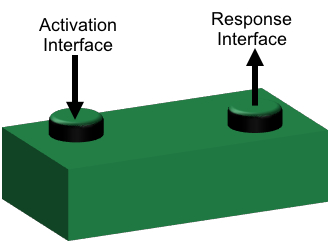}
\caption{Anatomy of a Resilient Design Pattern}
\label{Fig:PatternAnatomy}
\end{figure}

The basic template of a resilience design pattern is defined in an event-driven paradigm based on the insight that any resilience solution is necessary in the presence of, or sometimes in the anticipation of an anomalous event, such as a fault, error, or failure. The abstract resilience design pattern consists of a {\em behavior} and a set of {\em activation} and {\em response interfaces}. The appeal of defining the resilience design patterns in such an abstract manner is that they are universal. The abstract definition of the resilience design pattern behavior enables description of solutions that are free of implementation details. The instantiation of pattern behaviors may cover combinations of detection, containment and mitigation capabilities. The individual implementations of the same pattern may have different levels of performance, resilience, and power consumption. Also, the resilience pattern definition abstracts a pattern's interfaces from the implementation of these interfaces. 

\clearpage

\section{Classification of Resilience Design Patterns}
\label{sec:Resilience-Patterns-Classification}


\begin{figure}[tp]
  \centering
  \includegraphics[width=\textwidth]
                  {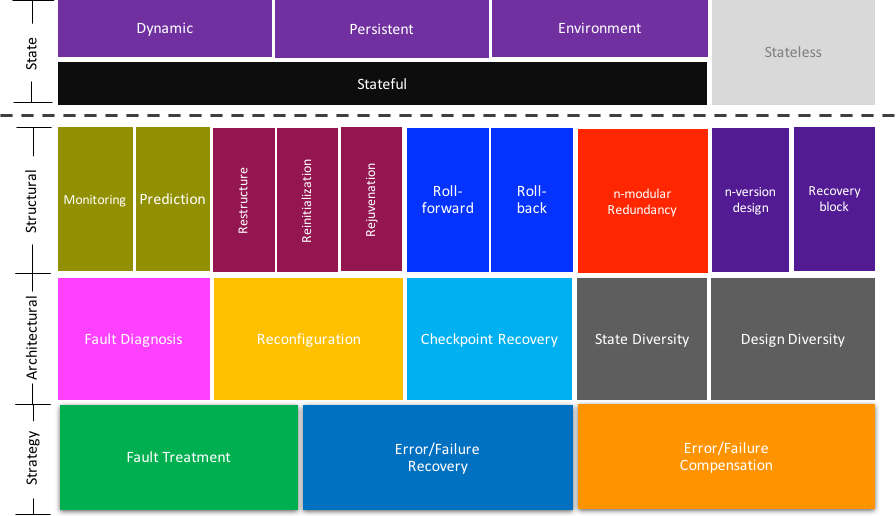}
  \caption{Classification of resilience design patterns}
  \label{Fig:PatternClassification}
\end{figure}

Architecting a HPC system and its software ecosystem is a complex process. In order to make the incorporation of resilience capabilities an essential part of the design process, the resilience design patterns are presented in a layered hierarchy. The hierarchy enables different stakeholders to reason about resilience capabilities based on their view of the system and their core expertise --- system architects may analyze protection coverage for the various hardware and software components that make up the system based on the patterns applied by each component; the designers of individual components may operate within a single layer of system abstraction and focus on instantiation of patterns based on local constraints and without the need to understand the overall system organization.

Resilience has two dimensions: (1) forward progress of the system and (2) data consistency in the system. Therefore, the design patterns are broadly classified into:

\begin{itemize}
  \item \textbf{State Patterns:}
        These patterns describe all aspects of the system structure that are
	relevant to the forward progress of the system. The correctness and 
	consistency of the system state ensures that the correct operation of 
	the system. The state implicitly defines the scope of the protection 
	domain that must be covered by a resilience mechanism. 
  \item \textbf{Behavior Patterns:}
        These design patterns identify common detection, containment, or
        mitigation actions that enable the components that realize these
        patterns to cope with the presence of a fault, error, or failure event.
\end{itemize}

This classification enables designers to separately reason about scope of the protection domain, and the semantics of the pattern behavior. The notion of state may be classified into three aspects \cite{Langou:2007}:
\begin{itemize}
\item \textit{Persistent/Static State}, which represents the data that is computed once in the initialization phase of the application and is unchanged thereafter.
\item \textit{Volatile/Dynamic State}, which includes all the system state whose value may change during the computation.
\item \textit{Operating Environment State}, which includes the data needed to perform the computation, i.e., the program code, environment variables, libraries, etc.
\end{itemize}

While certain behavior patterns may be applied to individual aspects of the system state, the state patterns may also be fused in order to enable the application of a single behavior pattern to more than one state pattern. Certain resilience behaviors may be applied without regard for state; such patterns are concerned with only the forward progress of the system.  The classification of state patterns also includes a \emph{stateless} pattern to enable designers to create solutions that define behavior without state.

The behavior patterns are presented in a layered hierarchy to provide designers with the flexibility to organize the patterns in well-defined and effective solutions:

\begin{itemize}
  \item \textbf{Strategy Patterns:}
        These patterns define high-level polices for a resilience solution.
        Their descriptions are deliberately abstract to enable hardware and 
        software architects to reason about the overall organization of
        the techniques used and their implications on the full system design.
  \item \textbf{Architecture Patterns:}
        These patterns convey specific methods of construction of a resilience
        solution. They explicitly convey the type of fault/error/failure event
        that they handle and provide detail about the key components and 
        connectors that make up the solution. 
  \item \textbf{Structure Patterns:}
        These patterns provide concrete descriptions of implementation
        strategies. They comprise of recipes that may be directly realized in
        hardware/software components. The implementation of these patterns is
        closely related to the state patterns.
\end{itemize}

{\bf Implementation patterns} bridge the gap between the design principles and the concrete details. These are compound patterns, i.e., patterns of patterns, and consist of a structure pattern and a state pattern. The specification also defines a {\em complete} resilience solution as one which provides detection/diagnosis of a fault, error or failure event, containment of its effects, and capabilities for mitigation to enable the affected system to continue intended operation.

\clearpage

\section{The Resilience Pattern Catalog}
\label{sec:Resilience-Pattern-Catalog}

\subsection{Describing Design Patterns}
Patterns are expressed in a written form in a highly structured format to enable HPC architects and designers to quickly discover whether the pattern is relevant to the problem being solved. For convenience and clarity, each resilience pattern follows the same prescribed format. There are three key reasons behind this pattern format: (1) to present the pattern solution in a manner that simplifies comparison of the capabilities of patterns and their use in developing complete resilience solutions, (2) to present the solution in a sufficiently abstract manner that designers may modify the solution depending on the context and other optimization parameters, and (3) to enable these patterns to be instantiated at different layers in the system.


\begin{mdframed}[backgroundcolor=white,linecolor=ornlGreen,linewidth=2pt,
                 nobreak=true]
\noindent\textbf{\emph{Name}}:\\
Identifies the pattern and provides a convenient way to refer to it, typically using a short phrase.

\vspace{1pt}

\noindent\textbf{\emph{Problem}}:\\
A description of the problem indicating the intent behind applying the pattern. This describes the intended goals and objectives that will accomplished with the use of this specific pattern.

\vspace{1pt}

\noindent\textbf{\emph{Context}}:\\
The preconditions under which the pattern is relevant, including a description of the system before the pattern is applied.

\vspace{1pt}

\noindent\textbf{\emph{Forces}}:\\
A description of the relevant forces and constraints, and how they interact or conflict with each other and with the intended goals and objectives. This description clarifies the intricacies of the problem and make explicit the trade-offs that must be considered.

\vspace{1pt}

\noindent\textbf{\emph{Solution}}:\\
A description of the solution that includes specifics of how to achieve the intended goals and objectives. 

\vspace{1pt}

\noindent\textbf{\emph{Capability}}:\\
The resilience management capabilities provided by this pattern, which may include detection, containment, mitigation, or a combination of these capabilities. 

\vspace{1pt}

\noindent\textbf{\emph{Protection Domain}}:\\
The resiliency behavior provided by the pattern extends over a certain scope, which may not always be explicit. Also, a solution may be suitable for a specific fault model. The description of scope and nature of fault model that is supported by the pattern enables designers to reason about the coverage scope in terms of the complete system.

\vspace{1pt}

\noindent\textbf{\emph{Resulting Context}}:\\
A brief description of the post-conditions arising from the application of the pattern. 

\vspace{1pt}

\noindent\textbf{\emph{Examples}}:\\
One or more sample applications of the pattern, which illustrate the use of the pattern for a specific problem, the context, and set of forces. This also includes a description of how the pattern is applied, and the resulting context.

\vspace{1pt}

\noindent\textbf{\emph{Rationale}}:\\
An explanation of the pattern as a whole, with an elaborate description of how the pattern actually works for specific situations. This provides insight into its internal workings of a resilience pattern, including details on how the pattern accomplishes the desired goals and objectives.

\vspace{1pt}

\noindent\textbf{\emph{Related Patterns}}:\\
The relationships between this pattern and other relevant patterns. These patterns may be predecessor or successor patterns in the hierarchical classification. The pattern may collaborate to complement, or enhance the resilience capabilities of other patterns. There may also be dependency relationships between patterns, which necessitate the use of co-dependent patterns in order to develop complete resilience solutions.

\vspace{1pt}

\noindent\textbf{\emph{Known Uses}}:\\
Known applications of the pattern in existing HPC systems, including any practical considerations and limitations that arise due to the use of the pattern at scale in production HPC environments.

\end{mdframed}
\clearpage

\textbf{Terminology}: The aim of defining a catalog of resilience design patterns is to provide reusable solutions to specific problems in a way that they may be instantiated in various ways, in hardware and software. Hardware design covers design of microarchitecture blocks, processor architecture, memory hierarchy design, network interface design, as well as design of racks, cabinet and system-level design. The scope of software design spans the spectrum of operating systems; runtimes for scheduling, memory management, communication frameworks, performance monitoring tools, computational libraries; compilers; programming languages; and application frameworks. In order to make the resilience pattern relevant to these diverse domains of computer system design, we describe solutions in a generic manner. The desrciptions use \textit{system} to refer to an entity that has the notion of structure and behavior. A \textit{subsystem} is a set of elements, which is a system itself, and is a component of a larger system, i.e., a system is composed of multiple sub-systems or components. For a HPC system architect, the scope of system may include 
compute nodes, I/O nodes, network interfaces, disks, etc., while an application developer may refer to a library interface, a function, or even a single variable as a system. The instantiation of the pattern descriptions interpret the notion of \textit{system} to refer to any of these hardware or software-level component. A \textit{full system} refers to the HPC system as a whole, and which is capable of running an application.   


%
%
\subsection{Strategy Patterns}

\subsubsection{Fault Treatment Pattern}
\par\textbf{\emph{Name}}: 
Fault Treatment Pattern

\par\textbf{\emph{Problem}}: 
The \texttt{Fault Treatment} pattern solves the problem of discovering and treating the presence of defects or anomalies in the system that have the potential to activate, leading to error or failure conditions in the system.   

\par\textbf{\emph{Context}}: 
The pattern applies to a system that has the following characteristics: 
\begin{itemize}
\item The system has well-defined parameters that enable a monitoring sub-system to discover the presence of anomalies in the behavior of a monitored sub-system. 
\item The interaction between the monitored and monitoring sub-systems is bounded in terms of time. 
\item The monitoring sub-system may have the capability modify the parameters of the monitored sub-system that enables the anomaly or defect to be removed before it results in an error or a failure. 
\end{itemize}

\par\textbf{\emph{Forces}}: 
\begin{itemize}
\item The interactions of the monitoring and monitored sub-systems may interfere with the operation of the monitored sub-system. The frequency of these interactions must be bounded.  
\item The time interval for the monitoring sub-system to gather data about the monitored sub-system and infer the presence of an anomaly or a defect incurs overhead to the operation of the monitored sub-system.  
\end{itemize}

\par\textbf{\emph{Solution}}: The \texttt{Fault Treatment} pattern provides a method that attempts to recognize the presence of an anomaly or a defect within a system, and creates conditions that prevents the activation of the fault into an error or failed state.

\par\textbf{\emph{Capability}}: 
The pattern provides fault mitigation semantics, which supports the following capabilities: 
\begin{itemize}
\item \textbf{Fault detection}, to detect anomalies during operation before they impact the correctness of the system state, and cause errors and failures.  
\item \textbf{Fault mitigation}, which includes methods to enable an imminent error or failure to be prevented, or the defect to be removed.
\end{itemize}
The \texttt{Fault Treatment} pattern may support either one, or both of these capabilities. 
 
\par\textbf{\emph{Protection Domain}}: 
The protection domain extends to the scope of the monitored sub-system, about which the monitoring system gathers data for discovering anomalies/defects. Since the pattern seeks to detect and alleviate a fault before activation, the protection domain implicitly extends to other sub-systems that are interfaced to the monitored sub-system. 

\par\textbf{\emph{Resulting Context}}: 
The \texttt{Fault Treatment} pattern requires the designer to identify parameters that indicate the presence of faults. The system design must include a monitoring sub-system, which introduces design complexity in the overall system design. When the monitoring sub-system is extrinsic to the monitored sub-system, the design effort may be simplified, and the interference between the sub-systems may be kept minimal. However, when the monitoring sub-system is intrinsic to the design of the monitored sub-system, the complexity of the design process increases, as well as the interaction between the sub-systems.  

\par\textbf{\emph{Examples}}:
Various hardware-based solutions for fault detection observe the attributes of a system, such as thermal state, timing violations in order to determine the presence of a defect in the behavior of the system that may potentially cause an error or failure. Similarly, software-based solutions detect the anomalies in the behavior of a system's data or control flow attributes to determine the presence of a fault.

\par\textbf{\emph{Rationale}}: The key benefit of incorporating fault mitigation patterns in a design, or deploying it during system operation is to preemptively recognize faults in the system, before they are activated and result in errors or partial/complete failures of the subsystem. The preventive actions avoid the need for expensive recovery and/or compensation actions.

\par\textbf{\emph{Related Patterns}}: The recovery and compensation patterns are complementary to the fault treatment patterns. Those patterns are necessary only the fault has been activated and an error or failure state exists in the system.

\par\textbf{\emph{Known Uses}}:
\begin{itemize}
\item Processor chips used in HPC systems contain thermal sensors that detect anomalous thermal conditions in the processor cores. When the temperature reaches a preset level, the sensor trips and processor execution is halted to prevent failure of the chip.
\item Software-based heartbeat monitoring for liveness checking of MPI processes enables detection of faults in a processor rank, before it may activate to result in failure of the MPI communicator.
\item The Cray RAS and Management System (CRMS) supports real-time fault monitoring of the status of Cray XT series system components, including the cabinets, blades, CPUs, Seastar processors and Seastar links.
\end{itemize}


\subsubsection{Recovery Pattern}
\par\textbf{\emph{Name}}: 
Recovery Pattern

\par\textbf{\emph{Problem}}: 
The \texttt{Recovery} pattern solves the problem of errors in the system, or failure of the system leading to catastrophic failure, which results in fail-stop behavior. In an HPC environment, the occurrence of errors or failures in the system results in catastrophic crashes, or incorrect results.      

\par\textbf{\emph{Context}}: 
The pattern applies to a system that is deterministic, i.e. forward progress of the system is defined in terms of the input state to the system and the execution steps completed since system initialization. The system must also have the following characteristics:  
\begin{itemize}
\item The error or failure in the system that the pattern handles must be detected; the pattern offers no implicit error/failure detection. 
\item The system has well-defined intervals that enables the pattern to transition the system to a known correct interval in response to an error/failure.  
\item The system is capable of compartmentalizing its state that is accurately representative of the progress of the system since initialization at the time such state is captured.  
\end{itemize}   

\par\textbf{\emph{Forces}}: 
\begin{itemize}
\item The pattern requires stable storage to capture system state, which increase overhead in terms of resources required by the system.
\item The process of compartmentalizing and capturing system state interferes with system operation. The error/failure-free overhead penalty must be minimized.  
\item The amount of state captured during each creation of a recovery point incurs space and time overheads.   
\item The frequency of creation of system state snapshots determines overhead: more frequent creation of recovery points increases system execution time, but reduces amount of lost work upon occurrence of an error/failure.    
\item The post-recovery state of the system must be as close as possible to an error/failure-free system state.
\item The interval between recovery from consecutive errors/failures must be less than the interval to create a stable recovery point from the present state of the system to enable system to make forward progress. 
\end{itemize}   

\par\textbf{\emph{Solution}}: 
The solution suggested by the \texttt{Recovery} pattern is based on the creation of snapshots of the system state and maintenance of these checkpoints on a stable storage system. Upon detection of an error or a failure, the checkpoints are used to recreate last known error/failure-free state of the system. Based on a temporal view of the system's progress, the error recovery may be either forward or backward.

\par\textbf{\emph{Capability}}: 
The \texttt{Recovery} pattern periodically preserves the essential aspects of the system state that may be subsequently used to resume operation from a known stable and correct version. The pattern handles an error or a failure by substituting an error-free state in place of the erroneous state. The pattern enables a system to tolerate errors/failures by resuming operation from a stable checkpointed version of the system that is free of the effects of the error/failure. The solution offered by this pattern is not dependent on the precise semantics of the error/failure propagation. The pattern does not offer error/failure detection capabilities.  

\par\textbf{\emph{Protection Domain}}: 
The protection domain for a \texttt{Recovery} pattern is determined by the extent of state that is captured during checkpoint operation, which accurately representatives the complete execution of the system. The broader the scope of the system state that is preserved, the larger is the scope of the system state that may be protected from an error/failure event.

\par\textbf{\emph{Resulting Context}}: 
With a recovery pattern, a system is capable of tolerating failures by substitution of erroneous/failed system state to a known previous stable state (backward recovery), or to an inferred future stable state (forward recovery). The frequency of creation of checkpoints determines the overhead to system operation; frequent checkpointing incurs proportionally greater overheads during error/failure-free operation, but reduces the amount of lost work when an error/failure event does occur. The latency of saving and restoring state influence the overhead during error/failure-free operation and the overhead of recovering from an error/failure respectively.

\par\textbf{\emph{Examples}}:
\begin{itemize}
\item Various checkpoint and restart libraries enable HPC applications to capture program state and commit the checkpoint files to parallel file systems. C/R capabilities in the OS such as BLCR \cite{BLCR:2002:LBNL} and libckpt \cite{Plank:1995:Libckpt} enables checkpointing the process state.  
\end{itemize}

\par\textbf{\emph{Rationale}}: 
Since the solution offered by this pattern is not dependent on either the type of error/failure, or the precise semantics of the error/failure propagation, the design effort and complexity in using this pattern in any system design in low. 

\par\textbf{\emph{Related Patterns}}: 
The \texttt{Compensation} pattern is complementary to the recovery pattern, although they both seek to create conditions to recreate correct state. The key difference between the \texttt{Recovery} and the \texttt{Compensation} patterns is the method used to maintain any additional state that is used for error/failure processing. While the compensation pattern use replication of the system, a recovery pattern relies on committing error/failure-free versions of the system to stable storage.

\par\textbf{\emph{Known Uses}}:
\begin{itemize}
\item The Berkeley Labs C/R library (BLCR) \cite{BLCR:2002:LBNL} is an extension to the Linux OS that supports creation checkpoint \& restart capabilities for Linux processes and also provides an interface for programmers to checkpoint application program state.
\end{itemize}

\subsubsection{Compensation Pattern}
\par\textbf{\emph{Name}}: 
Compensation Pattern

\par\textbf{\emph{Problem}}: 
The \texttt{Compensation} pattern solves the problem of errors in the system, or failure of the system leading to catastrophic failure, which results in fail-stop behavior. In an HPC environment, the occurrence of errors or failures in the system results in catastrophic crashes, or incorrect results.      

\par\textbf{\emph{Context}}: 
The pattern applies to a system that is deterministic, i.e. forward progress of the system is defined in terms of the input state to the system and the execution steps completed since system initialization. The system must also have the following characteristics:
\begin{itemize}
\item The error or failure in the system that the pattern handles must be detected, although the pattern may offer implicit error/failure detection. 
\item The error/failure must not be in the inputs provided to the system. 
\end{itemize}   

\par\textbf{\emph{Forces}}: 
\begin{itemize}
\item The pattern introduces penalty in terms of time (increase in execution time), or space (increase in resources required) independent of whether an errors or failure occurs. 
\item The error/failure-free overhead penalty must be minimized.  
\end{itemize}   

\par\textbf{\emph{Solution}}: 
The \texttt{Compensation} pattern is based on creation of a group of system replicas. The replicas are functionally identical and each replica receives identical inputs. 

\par\textbf{\emph{Capability}}: 
The pattern provides error/failure detection and correction, depending on the level of replication. The replicas of the system permit the system to continue operation despite the occurrence of a fail-stop failure by substituting a failed system replica with another. In order to recover from 2N failures in the system, there must be 2N + 1 distinct replicas. For the detection and correction errors, the outputs of the replicas of the system are compared by a monitor sub-system. When there are at least two replicas of the system, the monitor system compares the outputs to determine the presence of an error or failure in either replica. When the number of replicas is greater than two, and an odd number, the monitor performs majority voting on the outputs produced by the replicas, which enables incorrect outputs from replicas in error/failed state to be filtered out.  

\par\textbf{\emph{Protection Domain}}: 
The protection domain of the \texttt{Compensation} pattern extends to the scope of the system that is replicated. 

\par\textbf{\emph{Resulting Context}}: 
The pattern requires the replication of the system and its inputs. The design effort and complexity of replication of the system depends on the replication method. A naive replication requires low design effort; the design of functionally identical but independently designed versions of the replicas requires higher design effort. The preparedness of the replica during system operation to compensate for the error/failure state determines the level of overhead: The replica may \textit{active} or \textit{passive}; the replica state may be classified \emph{hot}, \emph{warm} and \emph{cold} are the forms of replication configurations based on the levels of intervention required for compensating for an error/failure. 

\par\textbf{\emph{Examples}}:
\begin{itemize}
\item Dual-modular redundancy for error detection; triple-modular redundancy for error detection and correction
\item Redundant information for compensation of data errors
\end{itemize}

\par\textbf{\emph{Rationale}}: 
The \texttt{Compensation} patterns enable systems to tolerate errors/failures by relying on the replicated versions of the system to substitute a failed system, or to infer and compensate for errors/failures by comparing the outputs of the replicas. 

\par\textbf{\emph{Related Patterns}}: 
The \texttt{Recovery} pattern is complementary to the compensation pattern, although they both seek to create conditions to recreate correct state. The key difference between the \texttt{Recovery} and the \texttt{Compensation} patterns is the method used to maintain any additional state that is used for error/failure processing. Unlike \texttt{Recovery} pattern, which uses a temporally forward or backward error/failure-free version of the system, a compensation pattern utilizes some form of redundancy to tolerate errors/failures in the system.

\par\textbf{\emph{Known Uses}}:
\begin{itemize}
\item Production HPC systems use memory modules that contain SECDED ECC, which maintain redundant bits per memory line. These redundant bits compensate for bit flip errors within the memory lines and enables detection and correction of certain errors.
\item The MR-MPI is an implementation of the MPI that transparently replicates and detects errors in MPI messages through active comparison between redundant execution instances of an application.
\end{itemize}

%
%
\subsection{Architectural Patterns}

\subsubsection{Fault Diagnosis Pattern}
\par\textbf{\emph{Pattern Name}}: Fault Diagnosis Pattern

\par\textbf{\emph{Problem}}: 
The \texttt{Fault Diagnosis} pattern solves the problem of identifying the presence of a defect or anomaly in the system. A fault in the system has the potential to activate, leading to the occurrence of an error or a failure. 

\par\textbf{\emph{Context}}: 
The pattern applies to a system that has the following characteristics:
\begin{itemize}
\item The system has well-defined parameters that enable a monitoring sub-system to discover the presence of anomalies in the behavior of a monitored sub-system.
\item The interaction between the monitored and monitoring sub-systems is bounded in terms of time.
\item The monitoring sub-system has the capability to analyze the behavior of the monitored sub-system.  
\end{itemize}

\par\textbf{\emph{Forces}}:
\begin{itemize}
\item The interactions of the monitoring and monitored sub-systems may interfere with the operation of the monitored sub-system. The frequency of these interactions must be bounded.
\item The time interval for the monitoring sub-system to gather data about the monitored sub-system and infer the presence of an anomaly or a defect incurs overhead to the operation of the monitored sub-system.
\end{itemize}

\par\textbf{\emph{Solution}}: 
The \texttt{Fault Diagnosis} pattern contains a monitoring sub-system that observes specific parameters of a monitored sub-system. The monitoring sub-system contains a range of acceptable values for the observed parameter to establish the notion of \textit{normal} operation of the monitored sub-system.  The monitoring sub-system observes deviations in the parameters to determine the presence of a fault, and the location of the fault.    

\par\textbf{\emph{Capability}}: 
The pattern provides a method that attempts to recognize the presence of an anomaly or a defect within a system and identifies the fault location and type. 

\par\textbf{\emph{Protection Domain}}:
The protection domain extends to the scope of the monitored sub-system, about which the monitoring system gathers data for discovering anomalies/defects. Since the pattern seeks to detect and alleviate a fault before activation, the protection domain implicitly extends to other sub-systems that are interfaced to the monitored sub-system.

\par\textbf{\emph{Resulting Context}}: 
The \texttt{Fault Diagnosis} pattern requires the designer to identify parameters that indicate the presence of faults. The system design must include a monitoring sub-system, which introduces design complexity in the overall system design. When the monitoring sub-system is extrinsic to the monitored sub-system, the design effort may be simplified, and the interference between the sub-systems may be kept minimal. However, when the monitoring sub-system is intrinsic to the design of the monitored sub-system, the complexity of the design process increases, as well as the interaction between the sub-systems.

\par\textbf{\emph{Rationale}}: 
The fault diagnosis patterns enable error/failure avoidance by detecting an anomaly in the system before it results in an error/failure.

\par\textbf{\emph{Examples}}:
Various hardware faults are detected through analysis of software symptoms, such as observation of latency of operations. 

\par\textbf{\emph{Related Patterns}}: 
In contrast to the error/detection recovery and compensation patterns, the fault diagnosis pattern is a passive pattern that observes system behavior and infers the presence of a fault based on the deviation from specified \textit{normal} behavior of the system.

\par\textbf{\emph{Known Uses}}:
\begin{itemize}
\item SMART (Self-Monitoring and Reporting Technology) is used in disk systems.
\end{itemize}

\subsubsection{Reconfiguration Pattern}
\par\textbf{\emph{Pattern Name}}: Reconfiguration Pattern

\par\textbf{\emph{Problem}}: 
The \texttt{Reconfiguration} pattern prevents a fault, error or failure in the system state affecting the correct operation of a system. 

\par\textbf{\emph{Context}}: 
The pattern applies to a system that has the following characteristics:
\begin{itemize}
\item The system that is deterministic, i.e. forward progress of the system is defined in terms of the input state to the system and the execution steps completed since system initialization.
\item The fault, error or failure in the system that the pattern handles must be detected; the pattern offers no implicit fault monitoring, prediction, or error/failure detection capability.
\item The system may be partitioned into N interconnected sub-systems. 
\end{itemize}

\par\textbf{\emph{Forces}}:
\begin{itemize}
\item The system must be able to be partitioned into a sub-set of sub-systems that is functionally equivalent to the fault, error, or failure-free version of the system.  
\item The reconfiguration may require the system to operate in degraded state using fewer than N sub-systems. The performance degradation of the system must be minimized.  
\end{itemize}

\par\textbf{\emph{Solution}}: 
The \texttt{Reconfiguration} pattern is based on isolation of the part of a system affected by a fault, error or failure, excluding the affected sub-system from interaction with other sub-systems, and restructuring the system to remain functionally equivalent to the system before the occurrence of the fault, error or failure event.  

\par\textbf{\emph{Capability}}: 
The pattern enables systems to tolerate the impact of a fault, error or failure by enabling the system to continue operation by preventing the affected part of the system from affecting the correctness of the system. 

\par\textbf{\emph{Protection Domain}}:
The protection domain of the \texttt{Reconfiguration} pattern spans the part of system whose state may be reconfigured, and yet is able to continue operating in a functionally equivalent operating state.  

\par\textbf{\emph{Resulting Context}}: 
\begin{itemize}
\item The reconfiguration of the system may result in the operation of the system in degraded condition. This incurs additional time overhead to the system. 
\item The pattern introduces additional design complexity since the system must remain functionally correct in multiple configurations. 
\end{itemize}

\par\textbf{\emph{Examples}}:
\begin{itemize}
\item Cluster management systems dynamically adapt the cluster configuration based on the health of various compute nodes in the system. 
\end{itemize}

\par\textbf{\emph{Rationale}}: 
The \texttt{Reconfiguration} pattern enables a system to tolerate to a fault, error or failure by adapting itself to the impact of the event and continuing to operate. The pattern enables systems to make forward progress by relying on the reconfigured version of the system. 

\par\textbf{\emph{Related Patterns}}: 
Like the \texttt{Checkpoint-Recovery} pattern, the \texttt{Reconfiguration} pattern is also supports recovery of the system from the impact of a fault, error or failure event. While the \texttt{Checkpoint-Recovery} pattern maintains snapshots of the system to stable storage to perform forward or backward recovery, the \texttt{Reconfiguration} pattern requires the system to adapt itself to isolate the impact of the event.    

\par\textbf{\emph{Known Uses}}:
The ULFM implementation of the MPI interface \cite{Bland:2013:IJHPCA} supports enables parallel applications to survive process failures during application execution by isolating the failed process, establishing agreement between the remaining processes in a MPI communicator and reconfiguring the communicator to include the remaining process ranks.

\subsubsection{Checkpoint Recovery Pattern}
\par\textbf{\emph{Pattern Name}}: Checkpoint Recovery Pattern

\par\textbf{\emph{Problem}}: 
The \texttt{Checkpoint-Recovery} pattern solves the problem of errors in the system, or failure of the system leading to catastrophic failure, which results in fail-stop behavior. 

\par\textbf{\emph{Context}}: 
The pattern applies to a system that is deterministic, i.e. forward progress of the system is defined in terms of the input state to the system and the execution steps completed since system initialization. The system must also have the following characteristics:
\begin{itemize}
\item The error or failure in the system that the pattern handles must be detected; the pattern offers no implicit error/failure detection.
\item The error or failure that the pattern handles must be transient, i.e., the error/failure must not repeatedly occur post-recovery  
\item The system has well-defined intervals that enables the pattern to transition the system to a known correct interval in response to an error/failure.
\item The system is capable of compartmentalizing its state that is accurately representative of the progress of the system since initialization at the time such state is captured.
\end{itemize}

\par\textbf{\emph{Forces}}:
\begin{itemize}
\item The pattern requires stable storage to capture system state, which increase overhead in terms of resources required by the system.
\item The process of compartmentalizing and capturing system state interferes with system operation. The error/failure-free overhead penalty must be minimized.
\item The amount of state captured during each creation of a recovery point incurs space and time overheads.
\item The frequency of creation of system state snapshots determines overhead: more frequent creation of recovery points increases system execution time, but reduces amount of lost work upon occurrence of an error/failure.
\item The post-recovery state of the system must be as close as possible to an error/failure-free system state.
\item The interval between recovery from consecutive errors/failures must be less than the interval to create a stable recovery point from the present state of the system to enable system to make forward progress.
\end{itemize}

\par\textbf{\emph{Solution}}: 
The solution suggested by the \texttt{Checkpoint-Recovery} pattern is based on the creation of snapshots of the system state and maintenance of these checkpoints on a stable storage system. Upon detection of an error or a failure, the checkpoints are used to recreate last known error/failure-free state of the system. Based on a temporal view of the system's progress, the error recovery may be either forward or backward.

\par\textbf{\emph{Capability}}: 
The \texttt{Checkpoint-Recovery} pattern periodically preserves the essential aspects of the system state that may be subsequently used to resume operation from a known stable and correct version. The pattern handles an error or a failure by substituting an error-free state in place of the erroneous state. The pattern enables a system to tolerate errors/failures by resuming operation from a stable checkpointed version of the system that is free of the effects of the error/failure. The solution offered by this pattern is not dependent on the precise semantics of the error/failure propagation. The pattern does not offer error/failure detection capabilities.

\par\textbf{\emph{Protection Domain}}:
The protection domain for a \texttt{Checkpoint-Recovery} pattern is determined by the extent of state that is captured during checkpoint operation, which accurately representatives the complete execution of the system. The broader the scope of the system state that is preserved, the larger is the scope of the system state that may be protected from an error/failure event.

\par\textbf{\emph{Resulting Context}}: 
With a recovery pattern, a system is capable of tolerating failures by substitution of erroneous/failed system state to a known previous stable state (backward recovery), or to an inferred future stable state (forward recovery). The frequency of creation of checkpoints determines the overhead to system operation; frequent checkpointing incurs proportionally greater overheads during error/failure-free operation, but reduces the amount of lost work when an error/failure event does occur. The latency of saving and restoring state influence the overhead during error/failure-free operation and the overhead of recovering from an error/failure respectively.

\par\textbf{\emph{Rationale}}: 
Since the solution offered by this pattern is not dependent on either the type of error/failure, or the precise semantics of the error/failure propagation, the design effort and complexity in using this pattern in any system design in low.

\par\textbf{\emph{Examples}}:
Application and system-level libraries that provide interfaces for creating checkpoints and restoring state are examples of the checkpoint-recovery pattern. 

\par\textbf{\emph{Related Patterns}}: 
The \texttt{State Diversity} and \texttt{Design Diversity} patterns are complementary to the \texttt{Checkpoint-Recovery} pattern, although they both seek to create conditions to recreate correct state. The key difference between these classes of patterns is the method used to maintain any additional state that is used for error/failure processing. While the diversity patterns use replication of the system, a recovery pattern relies on committing error/failure-free versions of the system to stable storage.

\par\textbf{\emph{Known Uses}}:
In order to tolerate fail-stop errors, current production-quality technologies rely on the classic rollback recovery approach using checkpoint restart (application-level or system-level) such BLCR, SCR.

\subsubsection{State Diversity Pattern}
\par\textbf{\emph{Pattern Name}}: State Diversity Pattern

\par\textbf{\emph{Problem}}:
The \texttt{State Diversity} pattern solves the problem of detecting and correcting errors or failures in the system state. 

\par\textbf{\emph{Context}}: 
The pattern applies to a system that has the following characteristics:
\begin{itemize}
\item The system must be deterministic, i.e. forward progress of the system is defined in terms of the input state to the system and the execution steps completed since system initialization. 
\item The cause of errors or failures experienced by the system may not be due to errors in the inputs.
\end{itemize}

\par\textbf{\emph{Forces}}:
\begin{itemize}
\item The pattern introduces penalty in terms of time (increase in execution time), or space (increase in resources required) independent of whether an errors or failure occurs.
\item The error/failure-free overhead penalty introduced by the replication of state must be minimized.
\end{itemize}

\par\textbf{\emph{Solution}}: 
The \texttt{State Diversity} pattern creates a group of N replicas of a system's state. The redundancy may include replication of the system's operation and/or the inputs to the system. Each of the N copies of the system state exist simultaneously. The redundant state versions of the systems are provided with the identical inputs, and their respective outputs are compared in order to detect and potentially correct the impact of an error or a failure in either replica of the systems. 

\par\textbf{\emph{Capability}}: 
The availability of replicated versions of the system state enable the following capabilities:
\begin{itemize}
\item Fail-over, which entails substitution of a replica in error or failed state with another replica that is error/failure-free.
\item Comparison, which entails observing the likeness of each replica's outputs as means to detect the presence of an error or failure in either replica.
\item Majority voting on the outputs produced by each replica system enables the detection of errors and failures, and filtering out the outputs that fall outside the majority.
\end{itemize}

\par\textbf{\emph{Protection Domain}}:
The protection domain of the pattern extends to the scope of the system state that is replicated. 

\par\textbf{\emph{Resulting Context}}: 
The design effort and complexity of replication of the system state requires low design effort since the replication entails creation of identical copies of the system state. 

\par\textbf{\emph{Rationale}}: 
The \texttt{State Diversity} patterns enable systems to tolerate errors/failures by relying on the replicated versions of the system state to substitute a failed system, or to infer and compensate for errors/failures by comparing the outputs of the replicas.

\par\textbf{\emph{Examples}}:
\begin{itemize}
\item Dual-modular redundancy for error detection; triple-modular redundancy for error detection and correction
\item Redundant information for compensation of data errors
\end{itemize}

\par\textbf{\emph{Related Patterns}}: 
The \texttt{State Diversity} and \texttt{Design Diversity} patterns are based on inclusion of redundancy in order to compensate for errors or failures. The diversity in the \texttt{State Diversity} pattern stems from the replication of the system's state unlike the \texttt{Design Diversity} pattern, which uses independently implemented versions of the system's design to tolerate errors or failures.   

\par\textbf{\emph{Known Uses}}:
The use of ECC memory and Chipkill in production HPC systems are known uses of the \texttt{State Diversity} pattern.

\subsubsection{Design Diversity Pattern}
\par\textbf{\emph{Pattern Name}}: Design Diversity Pattern

\par\textbf{\emph{Problem}}: 
The \texttt{Design Diversity} pattern solves the problem of detecting and correcting errors or failures in the behavior of the system that may occur due design faults in the system. 

\par\textbf{\emph{Context}}: 
The pattern applies to a system that has the following characteristics:
\begin{itemize}
\item The system has a well-defined specification for which multiple implementation variants may be designed. 
\item There is an implicit assumption of independence of between multiple variants of the implementation. 
\item The cause of errors or failures experienced by the system may not be due to errors in the inputs.  
\end{itemize}

\par\textbf{\emph{Forces}}:
\begin{itemize}
\item The pattern requires distinct implementations of the same design specification, which are created by different individuals or teams. 
\item The pattern increases the system complexity due to the need additional design and verification effort required to create multiple implementations.   
\item The error/failure-free overhead penalty due to disparity in the implementation variants must be minimized.
\end{itemize}

\par\textbf{\emph{Solution}}: 
The pattern enables systems to tolerate design faults in that may arise out of and incorrect interpretation of the specifications, or due to mistakes during implementation. The design diversity pattern entails partitioning the system into N replica sub-systems that are variants of a system design. These replicas are developed separately but are designed to a common specification. These design variants are applied in a time or space redundant manner. The redundant systems are provided with the identical inputs and their respective outputs are compared in order to detect and potentially correct the impact of an error or a failure in either replica of the systems. 

\par\textbf{\emph{Capability}}: 
The pattern relies on independently created, but functionally equivalent sub-system versions of a system specification. Since the sub-systems operate in parallel, in a time or space redundant manner, they are able to account for errors or failures caused by design flaws. 

\par\textbf{\emph{Protection Domain}}:
The protection domain extends to the scope of the system that is described by the design specification, of which multiple implementation variants are created.

\par\textbf{\emph{Resulting Context}}: 
Although the primary intent of the \texttt{Design Diversity} pattern is to enable systems to tolerate errors and failures due to design faults, the pattern also supports resilience transient errors/failures. Since the replica sub-systems are functionally identical and are deployed in a redundant manner, the likelihood that each replica is affected by the same transient error or failure is small.  

\par\textbf{\emph{Rationale}}: 
The intent behind applying this pattern is to eliminate the impact of human error during the implementation of a system. Due the low likelihood that different individuals or teams introduce identical bugs in their respective implementations, the pattern enables compensating for errors or failures caused by a bug in any one implementation of the same design. 

\par\textbf{\emph{Examples}}:
The concept of N-modular (NMR) programming is used in developing software development in which variants of a software developed by different teams, but to a common specification. These implementations may be applied in a time or space redundant manner to detect and correct errors or failures that are caused by bugs in the implementation.   

\par\textbf{\emph{Related Patterns}}: 
The \texttt{Design Diversity} and \texttt{State Diversity} patterns are based on inclusion of redundancy in order to compensate for errors or failures. In contrast to the \texttt{State Diversity} pattern, which replicates the system state, the \texttt{Design Diversity} pattern typically uses multiple versions of the system that are only functionally equivalent. 

\par\textbf{\emph{Known Uses}}:
Applications that require high precision floating point arithmetic, particularly application that require multiple-precision floating-point computations use multiple alternative compiler toolchains and library implementations that are functionally equivalent to ensure high precision in the computations.

%
%
\subsection{Structural Patterns}

\subsubsection{Monitoring Pattern}
\par\textbf{\emph{Pattern Name}}: Monitoring Pattern

\par\textbf{\emph{Problem}}: 
The \texttt{Monitoring} pattern solves the problem of analyzing the behavior of a system that indicates the immediate presence of a defect or anomaly in the system that has the potential to cause errors or failures in the system. 

\par\textbf{\emph{Context}}: The pattern applies to a system that has the following characteristics:
\begin{itemize}
\item The system has well-defined parameters that enable a monitoring sub-system to discover the presence of anomalies in the behavior of a monitored sub-system.
\item The interaction between the monitored and monitoring sub-systems is bounded in terms of time.
\item The monitoring sub-system has the capability to readily analyze the behavior of the monitored sub-system in order to identify anomalous behavior.
\end{itemize}

\par\textbf{\emph{Forces}}:
\begin{itemize}
\item The interactions of the monitoring and monitored sub-systems may interfere with the operation of the monitored sub-system. The frequency of these interactions must be bounded.
\item The time interval for the monitoring sub-system to gather data about the monitored sub-system and infer the presence of an anomaly or a defect incurs overhead to the operation of the monitored sub-system.
\end{itemize}

\par\textbf{\emph{Solution}}: 
The \texttt{Monitoring} pattern contains a monitoring sub-system that observes specific parameters of a monitored sub-system. The monitoring sub-system contains a range of acceptable values for the observed parameter to establish the notion of \textit{normal} operation of the monitored sub-system. The monitoring sub-system observes deviations in the parameters and the location of the anomaly to determine the root cause, type and precise location of a fault. 

\par\textbf{\emph{Capability}}: 
The pattern provides a method that attempts to recognize the presence of an anomaly or a defect within a system and identifies the fault location and type. 
\par\textbf{\emph{Protection Domain}}:
The protection domain extends to the scope of the monitored sub-system, about which the monitoring system gathers data for discovering anomalies/defects. Since the pattern seeks to detect and alleviate a fault before activation, the protection domain implicitly extends to other sub-systems that are interfaced to the monitored sub-system.

\par\textbf{\emph{Resulting Context}}: 
The \texttt{Monitoring} pattern requires the designer to identify parameters that indicate the presence of faults. The system design must include a monitoring sub-system, which introduces design complexity in the overall system design. When the monitoring sub-system is extrinsic to the monitored sub-system, the design effort may be simplified, and the interference between the sub-systems may be kept minimal. However, when the monitoring sub-system is intrinsic to the design of the monitored sub-system, the complexity of the design process increases, as well as the interaction between the sub-systems.

\par\textbf{\emph{Rationale}}:
The pattern enables the monitored sub-system to determine the presence of a fault and to analyze its root cause and location. The pattern enables the system to take precise corrective actions to prevent the activation of the fault to cause an error or failure event in the system.
 
\par\textbf{\emph{Examples}}:
Various hardware and software systems provide the capabilities to observe system behavior for the purpose of inferring the presence of faults, such as online self-tests. 

\par\textbf{\emph{Related Patterns}}: 
The structure of the \texttt{Monitoring} pattern is closely related to the \texttt{Prediction} pattern since they both contain monitoring and monitored sub-system entities. The key difference between these patterns is the amount of temporal information used by the patterns to assess the presence of a defect or anomaly in the system. The \texttt{Monitoring} pattern uses presently observed system parameters in contrast to the \texttt{Prediction} pattern, which uses historical trend information to forecast future fault events. 

\par\textbf{\emph{Known Uses}}:
The Intelligent Platform Management Interface (IPMI) provides message-based interface to collect sensors readings for health monitoring, including the data on temperature, fan speed, and voltage for the purpose of monitoring the hardware components in the system.

\subsubsection{Prediction Pattern}
\par\textbf{\emph{Pattern Name}}: Prediction Pattern

\par\textbf{\emph{Problem}}: 
The \texttt{Prediction} pattern solves the problem of identifying patterns of behavior that indicate the potential for future errors or failures in the system.  

\par\textbf{\emph{Context}}: 
The pattern applies to a system that has the following characteristics:
\begin{itemize}
\item The system has well-defined parameters that enable a monitoring sub-system to discover the presence of anomalies in the behavior of a monitored sub-system.
\item The interaction between the monitored and monitoring sub-systems is bounded in terms of time.
\item The monitoring sub-system has the capability to store historical data about the behavior of the monitored sub-system in order to analyze trends in fault occurrences.
\end{itemize}

\par\textbf{\emph{Forces}}:
\begin{itemize}
\item The interactions of the monitoring and monitored sub-systems may interfere with the operation of the monitored sub-system. The frequency of these interactions must be bounded.
\item The time interval for the monitoring sub-system to gather data about the monitored sub-system and infer the presence of an anomaly or a defect incurs overhead to the operation of the monitored sub-system.
\end{itemize}

\par\textbf{\emph{Solution}}: 
The \texttt{Prediction} pattern contains a monitoring sub-system that observes specific parameters of a monitored sub-system. The monitoring sub-system contains storage of the history of observed parameter values and fault events in the monitored sub-system. The monitoring sub-system uses past experiences of correlating the parameter values and fault occurrences to establish trends. These trends are used to predict occurrence of future faults based on observed deviations in the parameters.  
    
\par\textbf{\emph{Capability}}: 
The pattern provides a method that anticipates the occurrence of fault events based on patterns of behavior of the monitored sub-system that attempts to recognize the potential for future occurrences of an anomaly or a defect within a system. 

\par\textbf{\emph{Protection Domain}}:
The protection domain extends to the scope of the monitored sub-system, about which the monitoring system gathers data for discovering anomalies/defects. Since the pattern seeks to detect and alleviate a fault before activation, the protection domain implicitly extends to other sub-systems that are interfaced to the monitored sub-system.

\par\textbf{\emph{Resulting Context}}: 
The \texttt{Prediction} pattern requires the designer to identify parameters that may be used to forecast the occurrence of fault events. The system design must include a monitoring sub-system, which introduces design complexity in the overall system design. When the monitoring sub-system is extrinsic to the monitored sub-system, the design effort may be simplified, and the interference between the sub-systems may be kept minimal. However, when the monitoring sub-system is intrinsic to the design of the monitored sub-system, the complexity of the design process increases, as well as the interaction between the sub-systems. 

\par\textbf{\emph{Rationale}}: 
The pattern enables the monitored sub-system to use historical trends in system behavior before and during fault events to predict future fault events. If future fault events are predicted with high precision, then avoidance or preventive actions may be used. 

\par\textbf{\emph{Examples}}:
Hardware and software systems use correlations between past behaviors to predict the future occurrences of fault events, such as a memory device tends to show, for a given address, multiple repetitive correctable errors before showing an uncorrectable error.

\par\textbf{\emph{Related Patterns}}: 
The structure of the \texttt{Prediction} pattern is closely related to the \texttt{Monitoring} pattern since they both contain monitoring and monitored sub-system entities. The key difference between these patterns is the amount of temporal information used by the patterns to assess the presence of a defect or anomaly in the system. The \texttt{Prediction} pattern uses historical trend information to forecast future fault events in contrast to the \texttt{Monitoring} pattern, which uses presently observed system parameters.   
 
\par\textbf{\emph{Known Uses}}:
IPMI compliant servers have a System Event Log (SEL) which is a centralized, nonvolatile repository for all events generated. The trends in the SEL are used to make predictions of future events in the system.

\subsubsection{Restructure Pattern}
\par\textbf{\emph{Pattern Name}}: Restructure Pattern

\par\textbf{\emph{Problem}}:
The \texttt{Restructure} pattern solves the problem of a fault, error, or failure event affecting the correct operation of a system.

\par\textbf{\emph{Context}}: 
The pattern applies to a system that has the following characteristics:
\begin{itemize}
\item The system that is deterministic, i.e. forward progress of the system is defined in terms of the input state to the system and the execution steps completed since system initialization.
\item The fault, error or failure in the system that the pattern handles must be detected; the pattern offers no implicit fault monitoring, prediction, or error/failure detection capability.
\item The system may be partitioned into N interconnected sub-systems.
\end{itemize}

\par\textbf{\emph{Forces}}:
\begin{itemize}
\item The system must be able to be partitioned into a sub-set of sub-systems that is functionally equivalent to the fault, error, or failure-free version of the system.
\item The reconfiguration may require the system to operate in degraded state using fewer than N sub-systems. The performance degradation of the system must be minimized.
\end{itemize}

\par\textbf{\emph{Solution}}: 
The \texttt{Restructure} pattern is based on modifying the configuration between the N interconnected sub-systems to isolate the sub-system affected by a fault, error or failure. This reconfiguration excludes the affected sub-system from interaction with other sub-systems. The restructured system includes N-1 sub-systems, and yet seeks to remain functionally equivalent to the system before the occurrence of the fault, error or failure event.

\par\textbf{\emph{Capability}}: 
The pattern enables systems to tolerate the impact of a fault, error or failure by enabling the system to continue operation by preventing the affected part of the system from affecting the correctness of the system.

\par\textbf{\emph{Protection Domain}}:
The protection domain of the \texttt{Restructure} pattern spans the part of system whose state may be reconfigured, and yet is able to continue operating in a functionally equivalent operating state.

\par\textbf{\emph{Resulting Context}}:
\begin{itemize}
\item The reconfiguration of the system may result in the operation of the system in degraded condition. This incurs additional time overhead to the system.
\item The pattern introduces additional design complexity since the system must remain functionally correct in multiple configurations. 
\end{itemize}

\par\textbf{\emph{Rationale}}: 
The \texttt{Restructure} pattern enables a system to tolerate to a fault, error or failure by adapting itself to the impact of the event and continuing to operate. The pattern enables systems to make forward progress by relying on the reconfigured version of the system.

\par\textbf{\emph{Examples}}:
Dynamic page retirement schemes are an example of the restructure pattern, in which pages that have an history of frequent memory errors are removed from the pool of available pages.

\par\textbf{\emph{Related Patterns}}: 
The remaining reconfiguration patterns - the rejuvenation and reinitialization patterns - are closely related since they all seek to isolate the error/failed state of the system and prevent it from affecting the remaining error/failure-free part of the system.

\par\textbf{\emph{Known Uses}}:
Chipkill memory provides the capabilities to reduces the amount of available system memory when a DIMM experiences an escalated sequence of ECC memory errors. The reconfiguration enables HPC system operation to continue with degraded performance until the defective memory module is replaced.

\subsubsection{Rejuvenation Pattern}
\par\textbf{\emph{Pattern Name}}: Rejuvenation Pattern

\par\textbf{\emph{Problem}}:
The \texttt{Rejuvenation} pattern solves the problem of a fault, error, or failure event affecting the correct operation of a system.

\par\textbf{\emph{Context}}:
The pattern applies to a system that has the following characteristics:
\begin{itemize}
\item The system that is deterministic, i.e. forward progress of the system is defined in terms of the input state to the system and the execution steps completed since system initialization.
\item The fault, error or failure in the system that the pattern handles must be detected; the pattern offers no implicit fault monitoring, prediction, or error/failure detection capability.
\item The system may be partitioned into N interconnected sub-systems.
\item The fault, error, or failure must not be persistent.
\end{itemize}

\par\textbf{\emph{Forces}}:
\begin{itemize}
\item The system must be able to be partitioned into a sub-set of sub-systems that is functionally equivalent to the fault, error, or failure-free version of the system.
\item The rejuvenation is often a slow process that requires substantial additional overhead to identify the part of the system affected by the fault, error or failure, and to selectively reinitialize the system, in addition to overhead incurred due to any lost work.
\end{itemize}

\par\textbf{\emph{Solution}}: 
The \texttt{Rejuvenation} pattern requires isolating the specific part of the system affected by an error/failure and restoring or recreating the affected state such that the system may resume normal operation.

\par\textbf{\emph{Capability}}: 
The pattern requires the system operation to be halted and identifying the part of the system affected by the error/failure. Only the affected part of the system is restored to ensure correct operation of the system.

\par\textbf{\emph{Protection Domain}}:
The protection domain of the \texttt{Restructure} pattern spans the part of system whose state may be reconfigured, and yet is able to continue operating in a functionally equivalent operating state.

\par\textbf{\emph{Resulting Context}}: 
\begin{itemize}
\item The rejuvenation of the system expects the result in the operation of the system in degraded condition. This incurs additional time overhead to the system.
\item The overhead in terms of time to identify the specific state affected by the fault, error or failure, and restore the it to known correct state may be considerable.  
\end{itemize}

\par\textbf{\emph{Rationale}}: 
The \texttt{Rejuvenation} pattern enables a system to tolerate to a fault, error or failure by restoring the affect part of the system to known state that will ensure correct operation. Such targeted recovery prevents complete reset, or restructuring the system, both of which carry considerable overhead to the system operation.  

\par\textbf{\emph{Examples}}:
The targeted recovery of data structures in system software, such as kernel modules, permits recovery without the need to reinitialize the complete system. 

\par\textbf{\emph{Related Patterns}}: 
The remaining reconfiguration patterns - the reinitialization and restructure patterns - are closely related since they all seek to isolate the error/failed state of the system and prevent it from affecting the remaining error/failure-free part of the system.

\par\textbf{\emph{Known Uses}}:
Algorithm-based recovery methods for data corruptions in data structures that are used in numerical analysis problems use interpolation of neighboring data values to rejuvenate a data structure in error state.

\subsubsection{Reinitialization Pattern}
\par\textbf{\emph{Pattern Name}}: Reinitialization Pattern

\par\textbf{\emph{Problem}}:
The \texttt{Reinitialization} pattern solves the problem of a fault, error, or failure event affecting the correct operation of a system.

\par\textbf{\emph{Context}}:
The pattern applies to a system that has the following characteristics:
\begin{itemize}
\item The system that is deterministic, i.e. forward progress of the system is defined in terms of the input state to the system and the execution steps completed since system initialization.
\item The fault, error or failure in the system that the pattern handles must be detected; the pattern offers no implicit fault monitoring, prediction, or error/failure detection capability.
\item The system may be partitioned into N interconnected sub-systems.
\item The fault, error, or failure must not be persistent.
\end{itemize}

\par\textbf{\emph{Forces}}:
\begin{itemize}
\item The system must be able to be partitioned into a sub-set of sub-systems that is functionally equivalent to the fault, error, or failure-free version of the system.
\item The reinitialization is often a slow process that requires substantial additional overhead to reinitialize the system, in addition to overhead incurred due to lost work.
\end{itemize}

\par\textbf{\emph{Solution}}: 
To recover the error/failure, the pattern restores the system to its initial state. This causes system operation to \textit{restart} and a pristine reset of state, which implicitly cleans up the effects of the error/failure.

\par\textbf{\emph{Capability}}: 
The \texttt{Reinitialization} pattern performs a reset of the system state to restore pristine state before system operation is resumed.  
\par\textbf{\emph{Protection Domain}}:
Since the reinitialization causes reset of the system state, the protection domain of the \texttt{Reinitialization} pattern spans the complete system. 

\par\textbf{\emph{Resulting Context}}: 
The restoral of the system state to the initial state causes lost work, but guarantees the impact of the fault/error/failure is completely removed before service is resumed. 

\par\textbf{\emph{Rationale}}: 
The \texttt{Reinitialization} pattern is applied in conditions in which the recovery from the fault/error/failure instance is deemed impossible, or excessively expensive in terms of overhead to performance.

\par\textbf{\emph{Examples}}:
A system reboot is an instance of the \texttt{Reinitialization} pattern.

\par\textbf{\emph{Related Patterns}}: 
The remaining reconfiguration patterns - the rejuvenation and restructure patterns - are closely related since they all seek to isolate the error/failed state of the system and prevent it from affecting the remaining error/failure-free part of the system.

\par\textbf{\emph{Known Uses}}:
Various cluster management software systems enable malfunctioning nodes in the cluster to be reset by initiating reboot sequence for a specific node without disrupting the remaining nodes in the system.

\subsubsection{Roll-back Pattern}
\par\textbf{\emph{Pattern Name}}: Roll-back Pattern

\par\textbf{\emph{Problem}}:
The \texttt{Roll-back} pattern solves the problem of errors in the system, or failure of the system leading to catastrophic failure, which results in fail-stop behavior.

\par\textbf{\emph{Context}}:
The pattern applies to a system that is deterministic, i.e. forward progress of the system is defined in terms of the input state to the system and the execution steps completed since system initialization. The system must also have the following characteristics:
\begin{itemize}
\item The error or failure in the system that the pattern handles must be detected; the pattern offers no implicit error/failure detection.
\item The error or failure that the pattern handles must be transient, i.e., the error/failure must not repeatedly occur post-recovery  
\item The system has well-defined intervals that enables the pattern to transition the system to a known correct interval in response to an error/failure.
\item The system is capable of compartmentalizing its state that is accurately representative of the progress of the system since initialization at the time such state is captured.
\end{itemize}

\par\textbf{\emph{Forces}}:
\begin{itemize}
\item The pattern requires stable storage to capture system state, which increase overhead in terms of resources required by the system.
\item The process of compartmentalizing and capturing system state interferes with system operation. The error/failure-free overhead penalty must be minimized.
\item The amount of state captured during each creation of a recovery point incurs space and time overheads.
\item The frequency of creation of system state snapshots determines overhead: more frequent creation of recovery points increases system execution time, but reduces amount of lost work upon occurrence of an error/failure.
\item The post-recovery state of the system must be as close as possible to an error/failure-free system state.
\item The interval between recovery from consecutive errors/failures must be less than the interval to create a stable recovery point from the present state of the system to enable system to make forward progress.
\end{itemize}

\par\textbf{\emph{Solution}}:
The solution suggested by the \texttt{Roll-back} pattern is based on the creation of snapshots of the system state and maintenance of these checkpoints on a stable storage system. Upon detection of an error or a failure, the checkpoints are used to recreate last known error/failure-free state of the system. Based on a temporal view of the system's progress, the error/failure recovery is backward, i.e., the restored system state is a previous correct state of the system.

\par\textbf{\emph{Capability}}: 
The \texttt{Roll-back} pattern enables the system to recovery and resume operation from the point of occurrence of an error or a failure. However, the recovery of the system state typically leverages previously captured checkpointed state and repeats operations from the last stable known system state.  

\par\textbf{\emph{Protection Domain}}:
The protection domain for a \texttt{Roll-back} pattern is determined by the extent of state that is captured during checkpoint operation, which accurately representatives the complete execution of the system. The broader the scope of the system state that is preserved, the larger is the scope of the system state that may be protected from an error/failure event.

\par\textbf{\emph{Resulting Context}}: 
The time overhead introduced by the application of the \texttt{Roll-back} pattern during error-free operation depends on the frequency of taking checkpoints. For recovery, the amount of work lost also correlates with the frequency of the checkpoint operations. The worst-case scenario for recovery using this pattern is rolling-back to the start-up of the system.

\par\textbf{\emph{Rationale}}: 
The solution offered by this pattern is not dependent on either the type of error/failure, or the precise semantics of the error/failure propagation, the design effort and complexity in using this pattern in any system design in low.

\par\textbf{\emph{Examples}}:
The classic rollback recovery approach is implemented using various application-level or system-level checkpoint and restart frameworks. 

\par\textbf{\emph{Related Patterns}}: 
The roll-forward pattern is closely related to the roll-back pattern. The key difference between the two patterns is the temporal relation between the recovered state and the error/failure state. In the roll-back pattern the recovered state is based on a previous stable version of the system state.

\par\textbf{\emph{Known Uses}}:
\begin{itemize}
\item System-level checkpoint used in production HPC systems uses BLCR. Recent research in this domain focuses on the integration of incremental checkpointing in BLCR
\end{itemize}

\subsubsection{Roll-forward Pattern}
\par\textbf{\emph{Pattern Name}}: Roll-forward Pattern

\par\textbf{\emph{Problem}}:
The \texttt{Roll-forward} pattern solves the problem of errors in the system, or failure of the system leading to catastrophic failure, which results in fail-stop behavior.

\par\textbf{\emph{Context}}:
The pattern applies to a system that is deterministic, i.e. forward progress of the system is defined in terms of the input state to the system and the execution steps completed since system initialization. The system must also have the following characteristics:
\begin{itemize}
\item The error or failure in the system that the pattern handles must be detected; the pattern offers no implicit error/failure detection.
\item The error or failure that the pattern handles must be transient, i.e., the error/failure must not repeatedly occur post-recovery
\item The system has well-defined intervals that enables the pattern to transition the system to a known correct interval in response to an error/failure.
\item The system is capable of compartmentalizing its state that is accurately representative of the progress of the system since initialization at the time such state is captured.
\item The time to recover from an error or a failure must be minimized.
\end{itemize}

\par\textbf{\emph{Forces}}:
\begin{itemize}
\item The pattern requires stable storage to capture system state, which increase overhead in terms of resources required by the system.
\item The process of compartmentalizing and capturing system state interferes with system operation. The error/failure-free overhead penalty must be minimized.
\item The amount of state captured during each creation of a recovery point incurs space and time overheads.
\item The frequency of creation of system state snapshots determines overhead: more frequent creation of recovery points increases system execution time, but reduces amount of lost work upon occurrence of an error/failure.
\item The post-recovery state of the system must be as close as possible to an error/failure-free system state.
\item The interval between recovery from consecutive errors/failures must be less than the interval to create a stable recovery point from the present state of the system to enable system to make forward progress.
\end{itemize}

\par\textbf{\emph{Solution}}:
The solution suggested by the \texttt{Roll-forward} pattern is based on the creation of snapshots of the system state and maintenance of these checkpoints on a stable storage system. Upon detection of an error or a failure, the checkpoints are used to create a new correct error/failure-free state of the system, which enables the system to move forward. 

\par\textbf{\emph{Capability}}:
The pattern enables forward recovery, which entails steps that may involve access to checkpoints from previous stable checkpoints or external state information to recover from the impact of the error/failure. However, based on a temporal view of the system's progress, the error/failure recovery is forward, i.e., the restored system state enables forward progress from the point of occurrence of the error/failure in the system.

\par\textbf{\emph{Protection Domain}}:
The protection domain for a \texttt{Roll-forward} pattern is determined by the extent of state that is captured during checkpoint operation, which accurately representatives the complete execution of the system. The broader the scope of the system state that is preserved, the larger is the scope of the system state that may be protected from an error/failure event.

\par\textbf{\emph{Resulting Context}}: 
The \texttt{Roll-forward} pattern enables the system to recovery and resume operation from the point of occurrence of an error or a failure. However, the recovery of the system state may leverage previously captured checkpointed state but does not require the system to repeat operations from the last stable checkpoint.  

\par\textbf{\emph{Rationale}}: 
The solution offered by this pattern is not dependent on either the type of error/failure, or the precise semantics of the error/failure propagation, the design effort and complexity in using this pattern in any system design in low.

\par\textbf{\emph{Examples}}:
Applications that contain some form of algorithmic fault tolerance are capable of forward recovery provided an error keeps the application process alive. The application is able to compensate for the presence of an error and resume operation without the need to roll-back execution.  

\par\textbf{\emph{Related Patterns}}: 
The roll-back pattern is closely related to the roll-forward pattern. The key difference between the two patterns is the temporal relation between the recovered state and the error/failure state. In the roll-forward pattern the recovered state is based on a stable version of the system state that represents forward progress from the point of error/failure occurrence.

\par\textbf{\emph{Known Uses}}:
\begin{itemize}
\item The fault-tolerant MPI (FT-MPI) library supports roll-forward recovery of MPI applications.   
\item The GVR system versioning of these distributed arrays for resilience through which application-specified forward error recovery is made possible.
\end{itemize}

\subsubsection{N-modular Redundancy Pattern}
\par\textbf{\emph{Pattern Name}}: N-modular Redundancy Pattern

\par\textbf{\emph{Problem}}:
The \texttt{N-modular Redundancy} pattern solves the problem of detecting and correcting errors or failures in the system state.

\par\textbf{\emph{Context}}:
The pattern applies to a system that has the following characteristics:
\begin{itemize}
\item The system must be deterministic, i.e. forward progress of the system is defined in terms of the input state to the system and the execution steps completed since system initialization.
\item The cause of errors or failures experienced by the system may not be due to errors in the inputs.
\end{itemize}

\par\textbf{\emph{Forces}}:
\begin{itemize}
\item The pattern introduces penalty in terms of time (increase in execution time), or space (increase in resources required) independent of whether an errors or failure occurs.
\item The error/failure-free overhead penalty introduced by the replication of state must be minimized.
\end{itemize}

\par\textbf{\emph{Solution}}:
The \texttt{N-modular redundancy} pattern creates a group of N replicas of a system's state. The redundancy may include replication of the system's operation and/or the inputs to the system. Each of the N copies of the system state exist simultaneously. The redundant state versions of the systems are provided with the identical inputs, and their respective outputs are compared in order to detect and potentially correct the impact of an error or a failure in either replica of the systems.

\par\textbf{\emph{Capability}}:
The availability of replicated versions of the system state enable the following capabilities:
\begin{itemize}
\item Fail-over, which entails substitution of a replica in error or failed state with another replica that is error/failure-free.
\item Comparison, which entails observing the likeness of each replica's outputs as means to detect the presence of an error or failure in either replica.
\item Majority voting on the outputs produced by each replica system enables the detection of errors and failures, and filtering out the outputs that fall outside the majority.
\end{itemize}

\par\textbf{\emph{Protection Domain}}:
The protection domain of the pattern extends to the scope of the system state that is replicated.

\par\textbf{\emph{Resulting Context}}:
The design effort and complexity of replication of the system state requires low design effort since the replication entails creation of identical copies of the system state.

\par\textbf{\emph{Rationale}}:
The \texttt{N-modular redundancy} patterns enable systems to tolerate errors/failures by relying on the replicated versions of the system state to substitute a failed system, or to infer and compensate for errors/failures by comparing the outputs of the replicas.

\par\textbf{\emph{Examples}}:
\begin{itemize}
\item Dual-modular redundancy for error detection; triple-modular redundancy for error detection and correction
\item Redundant information for compensation of data errors
\end{itemize}

\par\textbf{\emph{Related Patterns}}:
The \texttt{N-modular redundancy} and the \texttt{N-version} patterns are based on inclusion of redundancy in order to compensate for errors or failures. The diversity in the \texttt{N-modular} pattern stems from the replication of the system's state unlike the \texttt{N-version} pattern, which uses independently implemented versions of the system's design to tolerate errors or failures.

\par\textbf{\emph{Known Uses}}:
\begin{itemize}
\item Implementations of the MPI interface, such as rMPI, MR-MPI, RedMPI support various forms of n-modular redundancy through replication of processes, the messages between MPI processes  
\item Charm++ prototypes also offer process-level replication
\item Production HPC systems also contain built-in n-modular redundancy for critical components, such as power supply modules, fans, etc. 
\end{itemize}

\subsubsection{N-version Design Pattern}
\par\textbf{\emph{Pattern Name}}: N-version Design Pattern

\par\textbf{\emph{Problem}}:
The \texttt{N-version Design} pattern solves the problem of detecting and correcting errors or failures in the behavior of the system that may occur due design faults in the system.

\par\textbf{\emph{Context}}:
The pattern applies to a system that has the following characteristics:
\begin{itemize}
\item The system has a well-defined specification for which multiple implementation variants may be designed.
\item There is an implicit assumption of independence of between multiple variants of the implementation.
\item The cause of errors or failures experienced by the system may not be due to errors in the inputs.
\end{itemize}

\par\textbf{\emph{Forces}}:
\begin{itemize}
\item The pattern requires distinct implementations of the same design specification, which are created by different individuals or teams.
\item The pattern increases the system complexity due to the need additional design and verification effort required to create multiple implementations.
\item The error/failure-free overhead penalty due to disparity in the implementation variants must be minimized.
\end{itemize}

\par\textbf{\emph{Solution}}: 
The \texttt{N-version design} pattern enables dealing with errors or failures due to Bohrbugs, although Heisenbugs may also be treated using this design pattern. The pattern entails creations of N independent versions of the system that are functionally identical, but designed independently. A majority voting logic is used to compare the results produced by each design version.

\par\textbf{\emph{Capability}}: 
In this pattern, each of the N (N >= 2) versions of the designs are independently implemented, but the versions are functionally equivalent systems. The versions are operated independently and the critical aspects of the system state are compared in order to detect and correct errors/failures due to Bohrbugs or Heisenbugs.

\par\textbf{\emph{Protection Domain}}:
The protection domain extends to the scope of the system that is described by the design specification, of which multiple implementation variants are created.

\par\textbf{\emph{Resulting Context}}: 
The extent to which each of the n design versions are different affects the ability of the pattern to tolerate errors/failures in the system. The use of the n-version design pattern requires significant design overhead in implementing and testing independent versions of a specification. Differences in the design may cause differences in timing in generating output values for comparison and majority voting - these differences incur overhead to the overall system operation.

\par\textbf{\emph{Rationale}}: 
The intent behind applying this pattern is to eliminate the impact of human error during the implementation of a system. Due the low likelihood that different individuals or teams introduce identical bugs in their respective implementations, the pattern enables compensating for errors or failures caused by a bug in any one implementation of the same design.

\par\textbf{\emph{Examples}}:
Various versions of the same software are used for the detection of errors due to bugs in the implementation of either version.  

\par\textbf{\emph{Related Patterns}}: 
The pattern is similar to the n-modular redundancy pattern, which entails creating replica versions of the state associated with the pattern and accounting for the presence of errors/failures through majority voting. The key difference between the patterns is the independence of design between the replica versions of the system.

\par\textbf{\emph{Known Uses}}:
\begin{itemize}
\item The DIVA processor architecture includes an out-of-order core as well as a simple in-order pipelined core. The in-order pipeline is functionally equivalent to the primary processor core and is used to detect errors in the design of the out-of-order processor core. 
\end{itemize}

\subsubsection{Recovery Block Pattern}
\par\textbf{\emph{Pattern Name}}: Recovery Block Pattern

\par\textbf{\emph{Recovery Block Problem}}:
The \texttt{Recovery Block} pattern solves the problem of detecting and correcting errors or failures in the behavior of the system that may occur due design faults in the system.

\par\textbf{\emph{Context}}:
The pattern applies to a system that has the following characteristics:
\begin{itemize}
\item The system has a well-defined specification for which multiple implementation variants may be designed.
\item There is an implicit assumption of independence of between multiple variants of the implementation.
\item The cause of errors or failures experienced by the system may not be due to errors in the inputs.
\end{itemize}

\par\textbf{\emph{Forces}}:
\begin{itemize}
\item The pattern requires distinct implementations of the same design specification, which are created by different individuals or teams.
\item The pattern increases the system complexity due to the need additional design and verification effort required to create multiple implementations.
\item The error/failure-free overhead penalty due to disparity in the implementation variants must be minimized.
\end{itemize}

\par\textbf{\emph{Solution}}: 
The \texttt{Recovery block} pattern is a flavor of the N-version design pattern in which a recovery block is invoked when the result from the primary version of the system fails an acceptance test. The recovery block is another implementation version of the same design specification based on which the primary system is implemented. 

\par\textbf{\emph{Capability}}: 
With the use of the \texttt{Recovery block} pattern, the system is composed of functional blocks. Each block contains at least a primary design and exceptional case handler along with an adjudicator. If the adjudicator does not accept the results of the primary system, it invokes the exception handler subsystem. This indicates that the primary system could not perform the requested service operation. An acceptance test is used to test the validity of the result produced by the primary version. If the result from the primary version passes the acceptance test, this result is reported and execution stops. If, on the other hand, the result from the primary version fails the acceptance test, another version from among the multiple versions is invoked and the result produced is checked by the acceptance test.

\par\textbf{\emph{Protection Domain}}:
The protection domain extends to the scope of the system that is described by the design specification, of which the recovery block implementation variant is created.

\par\textbf{\emph{Resulting Context}}: 
The extent which the primary design and recovery block versions of the system specification are different affects the ability of the pattern to tolerate errors/failures in the system. The use of the \texttt{Recovery block} design pattern requires significant design overhead in implementing and testing independent versions of a specification. Differences in the design may cause differences in timing in generating output values for comparison and majority voting - these differences incur overhead to the overall system operation.

\par\textbf{\emph{Rationale}}: 
This pattern relies on multiple variants of a design which are functionally equivalent but designed independently. The secondary recovery block design is used to perform recovery, if the system implementation of the primary design produces an output that suggests the presence of an error/failure of the primary system. This determination is made by the adjudicator sub-system.

\par\textbf{\emph{Examples}}:
Various application-based fault tolerance methods include verification routines that check for the validity of a computation.

\par\textbf{\emph{Related Patterns}}: 
The significant differences in the recovery block approach from N-version programming are that only one version is executed at a time and the acceptability of results is decided by an adjudicator test rather than by majority voting.

\par\textbf{\emph{Known Uses}}:
\begin{itemize}
\item Containment Domains provide recovery blocks in order to recover from errors in the computation included within the domain in order to generate correct output values. 
\item The SwiFT library provides language based implementation of the recovery block for use in C language programs.
\end{itemize}

%
%
\subsection{State Patterns}
\subsubsection{Persistent State Pattern}
\par\textbf{\emph{Pattern Name}}: Persistent State Pattern

\par\textbf{\emph{Problem}}: 
The \texttt{Persistent State Pattern} solves the problem of the separating the part of the system state that remains unchanged for the entire duration of system operation. 

\par\textbf{\emph{Context}}: 
The pattern applies to the state of the system that has the following characteristics: 
\begin{itemize}
\item The overall state of the system is deterministic, i.e., the system output state is determined solely by the input state. 
\item The system state may be described in terms of state that remains invariant for the duration of the system operation and state that changes during system operation. 
\end{itemize}

\par\textbf{\emph{Forces}}:
\begin{itemize}
\item The state patterns present an application-centric view of systems. The precise definition of aspects of the system state are invariant and those that change depends on the layer of system abstraction. At the application-level, the separation of this state is straightforward. However, in the hardware and system software layers, distinguishing between these types of state is non-trivial. 
\end{itemize}

\par\textbf{\emph{Solution}}: 
The persistent state refers to all aspects of a system's state that is computed when the system is initialized, but is not modified during the system operation. From the perspective of an HPC application, the persistent state includes program instructions and variable state that is computed upon application initialization.  

\par\textbf{\emph{Capability}}: 
The correctness of the persistent state is essential to correct execution of a program. The presence of any errors in the persistent state may not necessarily lead to immediate catastrophic failure of an application program's execution, but might lead the program on divergent paths that cause a failure at a future point in the system's operation.

\par\textbf{\emph{Protection Domain}}:
The \texttt{Persistent State} pattern defines the scope of the application program state that is computed during initialization. 

\par\textbf{\emph{Resulting Context}}: 
The persistent state pattern defines the scope of the static program state. Such scope forms the protection domain for a resilience behavioral pattern.

\par\textbf{\emph{Examples}}:
Various algorithm-based fault tolerance methods leverage the property of invariance in the persistent state. These methods maintain redundant information about the application variables in the static state that enables recovery to their default data values at any time during application execution. 

\par\textbf{\emph{Rationale}}: 
The isolation of the state that is persistent throughout an application program execution is supported by this pattern. The state invariance feature provided by this pattern enables the use of resilience behavioral patterns that leverage this property to detect and recovery errors/failure of such state.

\par\textbf{\emph{Related Patterns}}:
Together with the \texttt{Dynamic State} pattern and \texttt{Environment State} pattern, the \texttt{Persistent State} pattern defines the overall state of a system.


\subsubsection{Dynamic State Pattern}
\par\textbf{\emph{Pattern Name}}: Dynamic State Pattern

\par\textbf{\emph{Problem}}: 
The \texttt{Dynamic State Pattern} solves the problem of the encapsulating the part of the system state that changes during system operation. 

\par\textbf{\emph{Context}}: 
The pattern applies to the state of the system that has the following characteristics: 
\begin{itemize}
\item The overall state of the system is deterministic, i.e., the system output state is determined solely by the input state. 
\item The system state may be described in terms of state that remains invariant for the duration of the system operation and state that changes during system operation. 
\end{itemize}

\par\textbf{\emph{Forces}}:
\begin{itemize}
\item The state patterns present an application-centric view of systems. The precise separation of aspects of the system state that will be invariant and those that change depends on the layer of system abstraction. At the application-level, the separation of this state is straightforward. However, in the hardware and system software layers, distinguishing between these types of state is non-trivial. 
\end{itemize}

\par\textbf{\emph{Solution}}: 
The \texttt{Dynamic State Pattern} encapsulates the system state that changes as the system makes forward progress.  

\par\textbf{\emph{Capability}}: 
The state refers to all aspects of the program state that continuously changes as an application program executes. This includes the data values that are computed during system operation, or those that enable forward progress of the system (control-flow variables). 

\par\textbf{\emph{Resulting Context}}: 
For the perspective of an HPC program, the encapsulation of the dynamic state for the purpose of defining its resilience behavior enables correct operation and forward progress of the system.   

\par\textbf{\emph{Examples}}:
Algorithm-based fault tolerance strategies that guarantee resilience of the dynamic state actively track changes to state. Redundancy methods maintain copies of the change to the dynamic state in order to recover the version that is impacted by an error or failure. 

\par\textbf{\emph{Rationale/Capability}}: 
The isolation of the dynamic state that is updated throughout an application program execution is supported by this pattern. The \textit{dynamic} feature of this state pattern implies that any errors/failure in such state amounts to lost work. However, this isolation of dynamic state enables the use of resilience behavioral patterns that leverage this property to recover the error/failure without the need to abort and restart an application program.

\par\textbf{\emph{Related Patterns}}:
Together with the \texttt{Dynamic State} pattern and \texttt{Environment State} pattern, the \texttt{Persistent State} pattern defines the overall state of a system.


\subsubsection{Environment State Pattern}
\par\textbf{\emph{Pattern Name}}: Environment State Pattern

\par\textbf{\emph{Problem}}: 
The \texttt{Environment State Pattern} solves the problem of encapsulating the system state that supports the operation of the system.  
\par\textbf{\emph{Context}}: 
The pattern applies to the state of the system that has the following characteristics: 
\begin{itemize}
\item The overall state of the system is deterministic, i.e., the system output state is determined solely by the input state. 
\item The system state may be described in terms of state relevant to the core function of the system, called the \textit{primary state} and the system state that supports its function, called the \textit{ecosystem}.  
\end{itemize}

\par\textbf{\emph{Forces}}:
\begin{itemize}
\item The state patterns present an application-centric view of systems. The aspects of the system state that serves the primary function of the system and those that form the ecosystem is easier due to the convention of defining system layers of abstractions and the availability of well-defined interfaces between the layers. These interfaces enable designers to distinguish between the system's primary state and the environment state.  
\end{itemize}

\par\textbf{\emph{Solution}}: 
The \texttt{Environment State} pattern defines state that a system relies on to access the system resources and services that enable the system to fulfill its function. The environment also facilitates and coordinates the operation of various sub-systems. 

\par\textbf{\emph{Capability}}:
An error/failure in the environment state is often immediately catastrophic to the operation of the primary system. The encapsulation of the environment state provided by this pattern enables the development of separate resilience strategies for the ecosystem. 

\par\textbf{\emph{Protection Domain}}:
The \texttt{Environment State} pattern defines the scope of the components in the ecosystem that support the operation of the primary system. For an HPC program, this scope includes productivity tools and libraries, a runtime system, the operating system, file systems, communication channels, etc.

\par\textbf{\emph{Resulting Context}}: 
The system does not normally have complete control over its environment, but may have partial control to affect the environment through well-defined interfaces. Any changes to the environment typically affect the system operating within the environment directly. The encapsulation of the environment enables the resilience behavior of the environment state to be reasoned about separately from the resilience behavior of the primary system state.  

\par\textbf{\emph{Examples}}:
Operating-system based resilience mechanisms focus on the correctness of the data structure state within the kernel without concern for the resilience features of the application program. 

\par\textbf{\emph{Rationale}}: 
The encapsulation of the supporting sub-systems from the primary system state enables the application of specific behavioral resilience patterns that are general-purpose, which do not rely on any feature of the primary system.

\par\textbf{\emph{Related Patterns}}:
Together with the \texttt{Persistent State} pattern and \texttt{Dynamic State} pattern, the \texttt{Environment State} pattern defines the overall state of a system. 


\subsubsection{Stateless Pattern}
\par\textbf{\emph{Pattern Name}}: Stateless Pattern

\par\textbf{\emph{Problem}}: 
The \texttt{Stateless Pattern} solves the problem of defining resilience strategies that are independent of state.  

\par\textbf{\emph{Context}}:
The pattern applies to the state of the system that has the following characteristics: 
\begin{itemize}
\item The overall state of the system is deterministic, i.e., the system output state is determined solely by the input state.
\item The behavior and the progress of the system is does not depend on specific parts of the system state. 
\end{itemize}

\par\textbf{\emph{Forces}}:
\begin{itemize}
\item The state patterns present an application-centric view of systems. The precise separation of aspects of the system state that will be invariant and those that change depends on the layer of system abstraction. At the application-level, the separation of this state is straightforward. However, in the hardware and system software layers, distinguishing between these types of state is non-trivial. 
\end{itemize}

\par\textbf{\emph{Solution}}: 
The \texttt{Stateless} pattern provides the notion of the \textit{null} state in order to define resilience solutions that are independent of system state. 

\par\textbf{\emph{Resulting Context}}: 
\begin{itemize}
\item The stateless pattern is utilized together with behavioral resilience patterns whose actions do not necessitate operating on any aspect of the application program state.
\item The behavioral pattern used to build a resilience solution using the stateless pattern must be capable to deal with any additional side-effects in the system.
\end{itemize}

\par\textbf{\emph{Examples}}:
The use of the \textit{transaction} model to provide resilient behavior is an example of the \texttt{Stateless} pattern. Transactions support execution of a sequence of operations that may complete as a unit, or fail; the notion of partial execution is not supported. While the transaction may entail performing computation on data variables, the resilience of the data is independently managed; the resilience solution may be defined with a \texttt{Stateless} pattern.  

\par\textbf{\emph{Rationale/Capability}}: 
The pattern is the equivalent of a \textit{null} pattern that enables resilience solutions to be constructed without the requirement for the behavioral patterns to operate on the program state.

\par\textbf{\emph{Related Patterns}}:
While the \texttt{Persistent State} pattern \texttt{Dynamic State} pattern, and the \texttt{Environment State} pattern defines the complement of the overall state of a system, the \texttt{Stateless} pattern offers the notion of \texttt{null} state.


\clearpage

\section{Building Resilience Solutions using Resilience Design Patterns}
\label{sec:Building-Resilience-Solutions}

\begin{figure}[htp]
\centering
\includegraphics[width=\textwidth]
                {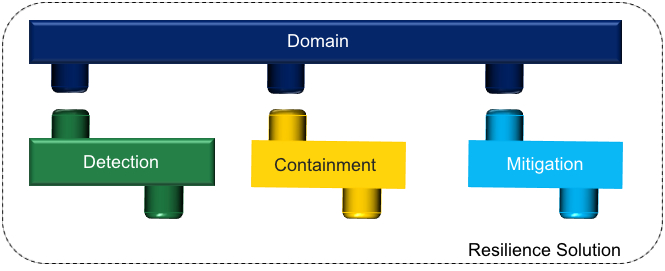}
\caption{Elements of a resilience solution for HPC systems and applications} 
\label{Fig:PatternSolutionElements}
\end{figure}

\subsection{Features of Resilience Solutions}
The resilience design patterns presented in the catalog offer solutions to problems that repeatedly occur in the design of resilience capabilities for HPC. Each pattern in the catalog presents a solution to a specific problem in detecting, recovering from, masking an error, or the scope of system state that is of interest to the resilience solution. These key constituents of a complete solutions are shown in Figure \ref{Fig:PatternSolutionElements}.  

The artifacts of a design process that uses the resilience design patterns are complete resilience solutions that provide fault/error/failure detection, containment and mitigation capabilities for a specific fault model. These solutions may be instantiated at multiple layers of system abstraction, and are relevant to various application and system scales. However, many of the patterns in the catalog individually provide partial solutions by supporting only one or two out of the detection, containment and mitigation solutions. For system and application designers to use these patterns in the construction of resilient versions of their designs, these patterns must be organized into a well-defined system of patterns.

A pattern framework enables the creation of the outline of the resilience solution that captures the dimensions and capabilities of the patterns, reveals and clarifies the relationships between the patterns. The combination of these patterns based on the guidelines offered by the hierarchical classification scheme enables the complete solutions for resilience to specific fault models in HPC systems. However, there is sufficient flexibility to adapt the solution to specific situations.  

\subsection{Design Spaces}
We define a framework that enables the composition of the resilience design patterns into practical solutions.

\begin{figure}[htp]
\centering
\includegraphics[width=0.25\textwidth]
                {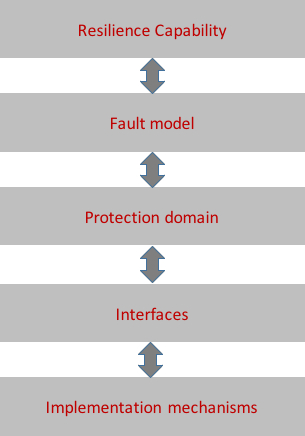}
\caption{Design Spaces for construction of resilience solutions using
         patterns}
\label{Fig:PatternDesignSpaces}
\end{figure}

In order to articulate a systematic method for customized designs, the framework is based on {\em design spaces} (Figure~\ref{Fig:PatternDesignSpaces}). These design spaces provide guidelines for the decision making in the design process, which consists of selection of the appropriate patterns based on the requirements of protection and the cost of using specific patterns.

\begin{itemize}
\item \textbf{Capability}:
      The patterns must support capabilities that enable the detection,
      containment, mitigation of faults/errors/failure events.
\item \textbf{Fault model}:
      The identification of the root causes of fault events and their impact
      and propagation through the system must be well-understood to provide
      effective solutions.
\item \textbf{Protection domain}:
      The definition of the protection domain enables clear encapsulation of
      the system scope over which the resilience patterns operate.
\item \textbf{Interfaces}:
      The identification and implementation of the activation and response
      interfaces for behavioral patterns affect the propagation of
      fault/error/failure event information.
\item \textbf{Implementation mechanisms}:
      The implementation design space is concerned with constraints imposed
      by specific features of hardware, execution or programming models,
      software ecosystems.
\end{itemize}

The structured design process enabled by these design spaces supports various approaches to create resilience solutions, including (i) a {\em top-down approach}; (ii) a {\em bottom-up approach}, as well as (iii) various {\em hybrid} approaches that enable designers to create solutions in the presence of practical constraints imposed by any hardware or software system features. Design spaces also provide a framework to guide the creation of cross-layered resilience solutions that leverage capabilities from multiple layers of the system abstraction. With the use of resilience patterns in the context of the framework provided by the design spaces, HPC system designers, users and application developers may evaluate the feasibility and effectiveness of novel resilience techniques, as well as analyze and evaluate existing solutions.

\clearpage

\section{Case Study: Checkpoint and Rollback}
\label{sec:CaseStudy-CR}

\begin{figure}[h]
\centering
\includegraphics[width=\textwidth]{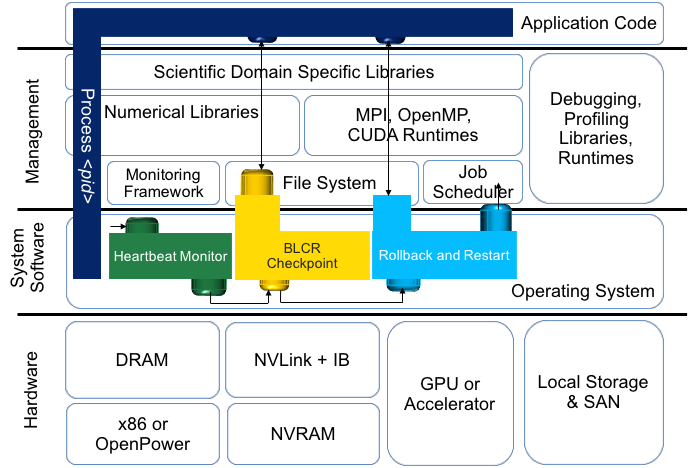}
\caption{Resilience Solution Case Study: Checkpoint \& Restart using BLCR} 
\label{Fig:PatternCaseStudy-CR}
\end{figure}

Checkpoint and restart (C/R) solutions are the most widely-used resilience solution in HPC systems. These solutions capture the image, or snapshot, of a running process and preserve it for later recovery. For parallel applications, the C/R framework's coordination protocols produce a global snapshot of the application by combining the state of all the processes in the parallel application. The checkpoint is typically committed to parallel file system on disk. Since C/R is a well-understood resilience strategy used in production HPC systems, the goal of this case study is to breakdown this solution and cast it within the resilience design patterns-based framework. The reexamination of this well-known resilience solution demonstrates the utility of the pattern-based framework in understanding the protection domain, capabilities, as well as the limitations of the C/R solution.   
 
For this case study, the resilience solution, which is illustrated in Figure \ref{Fig:PatternCaseStudy-CR}, is built using the BLCR (Berkley Lab's Checkpoint/Restart) \cite{BLCR:2002:LBNL} framework for a single process. In order to deconstruct this solution based on the structured pattern-based approach, we navigate the design spaces to methodically construct the solution. we focus on developing a complete resilience solution that enables systems to contend with single process failure. Since the fault model that our solution addresses is process failure, it is not necessary to identify the root cause of the fault and error that cause this failure. 

We identify patterns for:
\begin{itemize}
\item \textbf{\emph{Detection:}} For the detection of a process failure, we require an instantiation of the \texttt{Fault Treatment} pattern. Specifically, the solution requires \texttt{Fault Diagnosis} pattern to discover the location of the failure and the type of event, which is enabled by a \texttt{Monitoring} pattern. The instantiation of the \texttt{Monitoring} pattern is a kernel-level heartbeat monitor, which is deployed in the system to detects whether the process is alive.
 
\item \textbf{\emph{Containment:}} The BLCR framework provides containment by recovering a failed process from the last known stable process state from disk, which prevents the propagation of the failure.

\item \textbf{\emph{Recovery:}} BLCR also manages the recovery of the failure by instantiating the \texttt{Recovery} pattern, specifically the \texttt{Roll-back} structure pattern, whose architecture is framed using the \texttt{Checkpoint-Recovery} pattern. 

\item \textbf{\emph{Domain:}} Since the solution aims to provide resilience capabilities for the complete process, the solution fuses the \texttt{Persistent} and \texttt{Dynamic} state patterns. Therefore, the protection domain associated with the system-level checkpointing solution extends to the entire memory associated with a process. 

\end{itemize}

BLCR provides a completely transparent checkpoint of the process, which saves the current state of a process. The framework uses a coarse-grain locking mechanism to interrupt momentarily the execution of all the threads of the process, giving them a global view of its current state, and reducing the problem of saving the process state to a sequential problem. Since the entire state is saved (from CPU registers to the virtual memory map), the function call stack is saved. From the perspective of an application programmer, the checkpoint routine returns with a different error code, to let the caller know if this calls returns from a successful checkpoint or from a successful restart. The recovery after the detection of a process failure by the instantiation of the \texttt{Monitoring} pattern (the heartbeart monitor), entails restoring the checkpoint from the parallel file system on the same hardware, with the same software environment. 

\clearpage

\section{Case Study: Proactive Process Migration}
\label{sec:CaseStudy-ProcessMigration}

\begin{figure}[h]
\centering
\includegraphics[width=\textwidth]{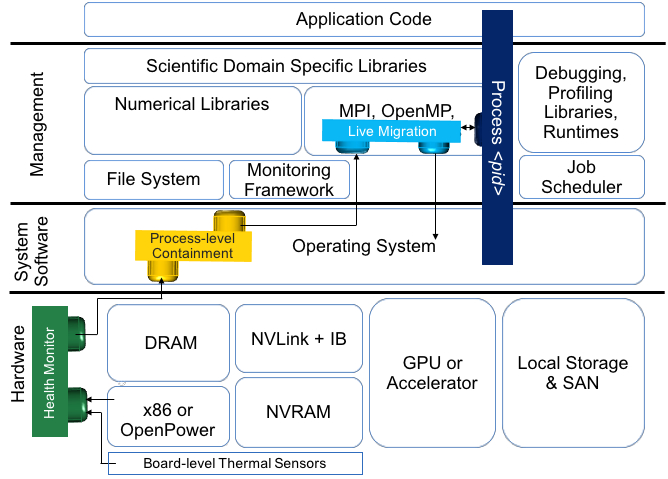}
\caption{Resilience Solution Case Study: Process Migration} 
\label{Fig:PatternCaseStudy-Migration}
\end{figure}

Various resilience strategies are inherently reactive, i.e., they respond to the occurrence of a fault, error or failure event and seek to prevent the event from affecting the correct execution of an HPC application. In this case study, we evaluate a proactive resilience solution using the resilience design pattern framework that enables a system to anticipate failures and provides failure avoidance capability through a process-level live migration mechanism \cite{Wang:2008}. The solution uses a combination of hardware and system software to handle resiliency in a manner that is transparent to the application developer. Also, the solution requires no changes in the application codes.

For the systematic study of the construction of the complete resilience solution based on the structured pattern-based approach, we address the specific design constraints that are imposed within each of the design spaces. For this case study, our emphasis is on the development of a complete resilience solution that avoids single process failure. We do not concern ourselves with the root cause of the fault and error that may cause the failure. 

We identify patterns for:
\begin{itemize}
\item \textbf{\emph{Detection:}} In order to proactively anticipate the occurrence of a failure, the solution must observe critical indicators that will predict the likelihood of a failure. We require the \texttt{Fault Treatment} strategy pattern, which must be instantiated as a \texttt{Fault Diagnosis} pattern within the system architecture. Since the detection entails anticipation of a future failure event, the structure pattern selected is the \texttt{Prediction} pattern, whose implementation requires reading board-level thermal sensors for health monitoring for each of the processors on the compute node.     

\item \textbf{\emph{Containment:}} A kernel level module provides containment for the fault by identifying the process that is executing on the CPU for which the \texttt{Prediction} pattern has assessed as vulnerable to a failure. 

\item \textbf{\emph{Recovery:}} The live migration is a kernel level module that is integrated with the MPI execution environment to support parallel applications. The \texttt{Recovery} strategy pattern is used by this solution. The architecture of the system must instantiate the \texttt{Reconfiguration} pattern and specifically the \texttt{Restructure} structural pattern in order to isolate the processor on which failure is predicted by the \texttt{Prediction} pattern and move the process to an alternative processor.  

\item \textbf{\emph{Domain:}} Since the solution aims to provide resilience capabilities for the complete process, the solution fuses the \texttt{Persistent} and \texttt{Dynamic} state patterns. Therefore, the protection domain associated with the system-level checkpointing solution extends to the entire memory associated with a process.
\end{itemize}

The \texttt{Prediction} pattern instantiation uses the standardized Intelligent Platform Management Interface (IPMI) which provides message-based interface to collect sensors readings for health monitoring, including the data on temperature, fan speed, and voltage. The instantiation of the \texttt{Restructure} pattern gathers the sensor data, and when the sensor reading exceeds a threshold value, the scheduler determines the availability of new destination nodes to complete the restructure of the job. If no spare nodes are available, the scheduler selects a compute node with lowest utilization. With the use of these pattern instantiations, this solution supports migration of a process as a precaution to potentially imminent failure by monitoring the health of each node. 

\clearpage

\section{Case Study: Cross-Layer Hardware/Software Hybrid Solution}
\label{sec:CaseStudy_ECC-ABFT}

\begin{figure}[h]
\centering
\includegraphics[width=\textwidth]{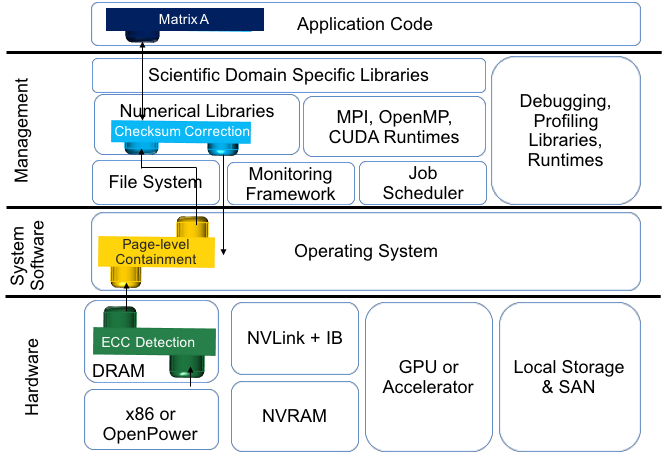}
\caption{Resilience Solution Case Study: Cross-Layer Design using ECC with ABFT} 
\label{Fig:PatternCaseStudy-CrossLayer}
\end{figure}

The pattern-based structured approach enables the design of resilience solutions that combine techniques across various layers of the system stack, which is referred to as cross-layer resilience. The aim of this case study is to use the framework of resilience design patterns to systematically explore various techniques available at multiple layers of the system stack and design a cross-layer combination that supports targeted, cost-effective resilience capabilities for a specific data structure within an application. The case study demonstrates practicality and effectiveness of our framework in developing novel resilience solutions.   
The proposed solution is intended to provide resilience capabilities for a matrix data structure in an application that uses a numerical method. The solution is designed to protect the matrix structure A from the impact of multi-bit corruptions due to errors in the DRAM memory.

We identify patterns for:
\begin{itemize}
\item \textbf{\emph{Detection:}} Since the memory is protected by error correcting codes (ECC), the solution leverages the hardware-based ECC. This is an instantiation of the \texttt{Compensation} pattern, and specifically the \texttt{State Diversity} pattern that supports the single error correction and double error detection capability.   

\item \textbf{\emph{Containment:}} The presence of double-bit error corruptions is detected by the \texttt{State Diversity} pattern and is communicated to the operating system via an interrupt mechanism. The containment is support by a module in the OS kernel that maps the physical address to the application address space, and notifies the library that provides the recovery pattern instantiation.  

\item \textbf{\emph{Recovery:}} The matrix A is protected by an algorithm-based fault tolerance (ABFT) method, specifically a checksum routine. This routine is an instantiation \texttt{Compensation} pattern, and specifically the \texttt{State Diversity} that is implemented within the numerical library.

\item \textbf{\emph{Domain:}} The solution is designed in order to guarantee the resilience of the data structure A, which as an operand matrix is an instantiation of the \texttt{Persistent} state pattern, since it is initialized during the initialization of the numerical method and does not change until the application converges.  
\end{itemize}

\clearpage

\section{Summary}
\label{sec:Summary}

The key to addressing the resilience challenge for future extreme-scale HPC systems make the design of resilience techniques an essential part of the system architecture and software development efforts. Towards the development of a systematic methodology for designing resilience solutions, design patterns provide reusable templates that may be used to build and refine resilience solutions. In this document we presented a set of design patterns that provide solutions to problems specific to the management of resilience in HPC systems. We identified and presented solutions that support detection, containment, recovery and masking in a structured design pattern format. We developed a classification scheme to enable designers to understand the capabilities of each pattern and the relationship between the various patterns in the pattern catalog. We developed a design framework to enable the composition and refinement of resilience solutions using the design patterns. The resilience design patterns and the design framework offer a systematic way to investigate the effectiveness and efficiency of a resilience solution. They also provide a structured approach for optimizing the trade-off, at design time or runtime, between the key system design factors: performance, resilience, and power consumption.

\clearpage

\bibliography{references}
\bibliographystyle{plain}
\end{document}